\documentclass[aps,prc,twocolumn,showpacs,floatfix,nofootinbib]{revtex4-1}
\usepackage[usenames,dvipsnames]{color}
\usepackage[normalem]{ulem}
\usepackage{comment}
\usepackage{textcomp}
\usepackage{amsmath,graphicx}

\begin{document}

\title{Thermal noise in non-boost-invariant dissipative hydrodynamics}
\author{Chandrodoy Chattopadhyay and Subrata Pal}
\affiliation{Department of Nuclear and Atomic Physics, Tata Institute of
Fundamental Research, Homi Bhabha Road, Mumbai 400005, India}

\begin{abstract}
We study the effects of hydrodynamic fluctuations in non-boost-invariant longitudinal expansion
of matter formed in relativistic heavy ion collisions. We formulate the theory of thermal noise
within second-order viscous hydrodynamics treating noise as a perturbation on top of the non-boost-invariant flow.
We develop a numerical simulation model to treat the (1+1)-dimension hydrodynamic evolution.
The code is tested to reproduce the analytic results for the Riemann solver for expansion of matter in vacuum.
For viscous hydrodynamic expansion, the initial energy density distribution are obtained by
reproducing the measured charged hadron rapidity distribution at the RHIC energies.
We show that the longitudinal rapidity correlations arising from space-time dependent thermal noise
and from an induced thermal perturbation have distinct structures.
In general, the rapidity correlations are found to be dominated by temperature fluctuations at
small rapidity separation and velocity fluctuations at large rapidities.
We demonstrate that thermal noise produce ridge-like two-particle rapidity correlations
which persist at moderately large rapidities. The magnitude and pattern of the correlations are quite
sensitive to various second-order dissipative formalisms and to the underlying equations of state,
especially at large rapidities. The short-range part of the rapidity correlation is found to be
somewhat enhanced as compared to that in boost-invariant flow of matter.
\end{abstract}

\pacs{25.75.Ld, 24.10.Nz, 47.75+f}


\maketitle

\section{Introduction}

Relativistic dissipative hydrodynamic has become the state-of-the-art model to study the evolution
of hot and dense matter formed at the Relativistic Heavy-ion Collider (RHIC)
\cite{Adams:2005dq,Adcox:2004mh} and at the Large Hadron Collider (LHC) 
\cite{ALICE:2011ab,ATLAS:2012at,Chatrchyan:2013kba}. Hydrodynamical model analysis of 
the large anisotropic flow observed in the plane transverse to the reaction plane has 
established the formation of a near-equilibrated strongly coupled Quark-Gluon-Plasma (QGP) 
with a small shear viscosity to entropy density ratio $\eta_v/s$. The flow is found to 
originate mostly during the initial stages of dynamical evolution 
\cite{Huovinen:2001cy,Heinz:2001xi}. Considerable efforts are
underway for an accurate determination of the transport properties of the QGP formed.

Inspite of the success of hydrodynamic models in the description of relativistic heavy-ion
collisions, considerably uncertainty prevails. This relates to the formulation of dissipative hydrodynamics,
the correct initial conditions, and the numerical implementation. At present the models
used are commonly based on approaches, such as the second-order dissipative (causal) 
equations in the M\"uller-Israel-Stewart (MIS) framework 
\cite{Muller:1967zza,Israel:1979wp,Muronga:2003ta,Romatschke:2007mq},
the Chapman-Enskog- (CE-) like iterative expansion of the Boltzmann equation in the 
relaxation-time approximation \cite{Jaiswal:2013npa,Bhalerao:2013pza,Chattopadhyay:2014lya},
the second-order viscous hydrodynamics from AdS/CFT correspondence,
and the anisotropic hydrodynamics \cite{Strickland:2014eua,Heinz:2014zha}.
For a reasonable description of the flow data, all these models require a very early thermalization 
proper time of $\tau \approx 0.2-0.6$ fm/c, that, hitherto, lacks a proper explanation.

One of the major uncertainties in the hydrodynamic model extraction of $\eta_v/s$ lies with
the initial state models. In fact, viscous hydrodynamic descriptions with different initial 
conditions, can be made compatible with the data for elliptic ($n=2$) and triangular ($n=3$) 
flow harmonics $v_n = \langle \cos(n-\Psi_n)\rangle$, for different 
tuned values of $\eta_v/s$ \cite{Luzum:2008cw,Schenke:2011bn,Qiu:2011hf}.
While the elliptic flow is driven primarily by the hydrodynamic
response of the initial overlap geometry of the colliding nuclei, the odd harmonics are 
solely governed by the initial-state fluctuations of the nucleon position in the nuclei
\cite{Alver:2010gr}. Pre-equilibrium parton dynamics and fluctuations in the parton 
production and scattering was shown to have a crucial effect on the final anisotropic 
flow \cite{Gale:2012rq,Bhalerao:2015iya,Chattopadhyay:2017bjs}. Further sources of fluctuations 
pertain to energy deposition (and its evolution) by a partonic jet 
propagating in the hydrodynamic medium \cite{Staig:2010pn} and the treatment of particilization 
of the fluid cells at freeze-out \cite{Tachibana:2017syd}.

In contrast, hydrodynamic fluctuations arising due to intrinsic thermal (and particle number)
fluctuations in each fluid cell occur during the entire evolution of the system
\cite{Kapusta:2011gt,Young:2013fka,Kapusta:2014dja,Young:2014pka,Albright:2015uua,Nagai:2016wyx}.
Based on the fluctuation-dissipation theorem, it is natural, that any dissipative system 
close to thermal equilibrium should
should exhibit thermal fluctuations. In heavy-ion collisions, as the transverse size
of the participant zone is about 5-10 fm, and the evolution stage lasts for about
$\tau \approx 10$ fm/c, thermal fluctuations in the fluid medium could have measurable and 
important consequences. The formulation of hydrodynamic fluctuations in the
nonrelativistic limit \cite{Landau} was recently extended to relativistic hydrodynamic regime. 
As an application of the stochastic thermal noise in relativistic heavy-ion collisions, 
it was demonstrated within boost-invariant one-dimensional (Bjorken) expansion of the fluid, 
that the two-particle rapidity correlations exhibit ridge-like structures observed 
in collisions at RHIC and LHC \cite{Kapusta:2011gt}. 

The thermal fluctuation of the energy-momentum tensor $\Xi^{\mu\nu}$ has a nontrivial autocorrelation
$\langle \Xi^{\mu\nu}(x) \Xi^{\alpha\beta}(x') \rangle \sim T \eta_v \: \delta^4(x-x')$ 
\cite{Kapusta:2011gt,Young:2013fka}.
Due to the Dirac-delta function, the energy and momentum density averaged value of this 
white noise becomes $\sim 1/\sqrt{\Delta V \Delta t}$. Thus even for small shear viscosities,
the white noise sets a lower limit on the system cell size $\Delta V$ that is essentially
comparable to the correlation length. Consequently, white noise could lead to large gradients 
which makes the basic hydrodynamic formulation (based on gradient expansion) questionable.
One possible way to overcome this is by using thermal fluctuations nonperturbatively via
colored noise \cite{Kapusta:2014dja}. Alternatively, white noise can be implemented 
by treating fluctuations as perturbations (in a linearized hydrodynamic framework) on top of 
a baseline non-fluctuating hydrodynamic evolution 
\cite{Kapusta:2011gt,Young:2013fka,Chattopadhyay:2017rgh}.
While analytic solution of hydrodynamic fluctuation exists within relativistic Navier-Stokes theory
(for a conformal fluid) with idealized boost-invariant dynamics in Bjorken flow \cite{Kapusta:2011gt}
and Gubser flow \cite{Yan:2015lfa}, numerical simulations of thermal fluctuation were performed 
for various second-order dissipative hydrodynamics for Bjorken flow profiles \cite{Chattopadhyay:2017rgh}.

It is however important to realize that the propagation of the fluctuations over large distances and times 
critically depend on the underlying expansion of the fluid. As compared to a static fluid
\cite{Landau}, the observables related to the fluctuations, (namely, the two-particle rapidity 
correlations and harmonic flow $v_n$ distributions), calculated at the freeze-out time (or temperature), 
would have different features. The thermal noise correlators have been calculated and 
their phenomenological applications for the boost-invariant (Bjorken) expansion of matter 
have been explored \cite{Kapusta:2011gt,Chattopadhyay:2017rgh}.
However, the longitudinal boost-invariant scenario could only give a reasonable description 
for the midrapidity region during the initial stages of relativistic heavy ion collisions 
\cite{Bjorken:1982qr}. At large rapidities and due to finite size of the expanding fluid, thermal
noise would exhibit a different behavior. As a matter of fact, even the noiseless non-boost-invariant
longitudinal expansion could create large velocity and energy gradients at the cell
boundaries near large space-time rapidities that may severely affect the baseline hydrodynamic 
evolution \cite{Ryblewski:2010bs,Martinez:2010sd,Florkowski:2016kjj}. 

In this paper, we formulate the hydrodynamic (thermal) fluctuation in the non-boost-invariant
(1+1)D longitudinal expansion of viscous matter within the M\"uller-Israel-Stewart 
and Chapman-Enskog theories.
For the perturbative application of the white noise, we have developed a (1+1)D viscous
hydrodynamic simulation in the Milne coordinates ($\eta, \tau$) that is suitable to
study relativistic heavy-ion collisions. The code has been tested for stability against
shock waves across the cell boundaries and at large rapidities by comparing with the
Riemann and Landau-Khalatnikov wave solutions. Within linearized second-order viscous
hydrodynamic framework, we treat the fluctuations as perturbation on top of the background
(1+1)D expanding viscous medium. We first consider a single thermal perturbation and explore
the resulting longitudinal rapidity-correlations to gain insight into the more complex case of
thermal noise generated at all space-time points. We then perform extensive numerical simulation
of hydrodynamic fluctuations and study the rapidity correlations for the commonly used hydrodynamics
dissipative theories, namely MIS and CE formalisms and for the conformal
and lattice QCD equation of state (EoS). In particular, we will show that distinct
magnitude and structures in the rapidity correlations are obtained for these different cases.

The paper is organized as follows. In Sec. II we formulate the theory of hydrodynamic
fluctuations in the linearized limit for non-boost-invariant hydrodynamic expansion of viscous fluid.
We derive analytical expressions for the two-particle rapidity correlations from thermal noise at
freeze-out. In Sec. III A we test the (1+1)D hydrodynamic code with simple wave solutions.
We then constrain in Sec. III B the initial conditions of the viscous hydrodynamic code from fits to
the measured rapidity distribution of hadrons in central Au+Au collisions at RHIC.
With these parameter sets we present results for various rapidity correlators from an
induced thermal fluctuation and thermal noise in Sec. III C and III D, respectively. 
In Sec. III E the results for rapidity correlations
due to thermal noise are compared for the MIS and CE viscous approaches and for various EoS.
A summary and conclusions are presented in Sec. IV.

\section{Thermal noise in relativistic hydrodynamics}

\subsection{Fluctuation-dissipation relations for causal second-order theories}

The hydrodynamic evolution of a system is governed by the conservation equations 
for particle current, $\partial_\mu N^\mu=0$, and the 
energy-momentum tensor, $\partial_\mu T^{\mu\nu} =0$, where 
\begin{align}\label{NTD:eq}
N^\mu &= n u^\mu + n^\mu,  \nonumber\\
T^{\mu\nu} &= \epsilon u^\mu u^\nu - (p+\Pi) \Delta^{\mu \nu} + \pi^{\mu\nu}.
\end{align}
Here $n$ is the number density, $\epsilon$ and $p$ are respectively the energy density 
and pressure in the fluid's local rest frame (LRF), $\pi^{\mu\nu}$ is the shear 
pressure tensor, $n^\mu$ is the particle diffusion current and $\Pi$ the local bulk viscous pressure.
$\Delta^{\mu\nu}=g^{\mu\nu}-u^\mu u^\nu$ is the projection operator on the 
three-space orthogonal to the hydrodynamic four-velocity $u^\mu$ in the LRF 
that is defined by the Landau-matching condition $T^{\mu\nu}u_{\nu} = \epsilon u^\mu$. 
We will disregard particle flow $N^\mu$, which is a reasonable 
approximation due to very small values of net-baryon number formed at RHIC and LHC. 

In the relativistic Navier-Stokes (first order) theory, the instantaneous constituent equations
for the bulk and shear pressures are
\begin{align}\label{NS:eq}
\Pi_{{NS}} = - \zeta \nabla \cdot u,  ~~~
\pi_{NS}^{\mu\nu}  = 2\eta_v \nabla^{\langle u}u^{\nu\rangle} ,
\end{align}
where the transport coefficients $\zeta, \eta_v \geq 0$ are the bulk and shear viscosity, and 
$\nabla^{\langle u}u^{\nu\rangle} = (\nabla^\mu u^\nu + \nabla^\nu u^\mu)/2 - 
(\nabla \cdot u) \Delta^{\mu\nu}/3$ and $\nabla^\mu = \Delta^{\mu\nu}\partial_\nu$.
In the MIS theory \cite{Muller:1967zza,Israel:1979wp,Muronga:2003ta,Romatschke:2007mq},
derived from positivity of entropy four-current divergence, 
the second-order dissipative hydrodynamic equations,
\begin{align}\label{IS:eq}
D\Pi &= - \frac{1}{\tau_\Pi} \left( \Pi  + \zeta \nabla \cdot u \right), \nonumber\\
\Delta^{\mu\nu}_{\alpha\beta}D\pi^{\alpha\beta} &\approx - \frac{1}{\tau_\pi} \left( \pi^{\mu\nu} 
- 2\eta_v \nabla^{\langle u}u^{\nu\rangle} \right) - \frac{4}{3} \pi^{\mu\nu} \theta ,
\end{align}
restore causality by enforcing the bulk and shear pressures to relax to their first-order values via the 
relaxation times $\tau_\Pi = \zeta \beta_0$ and $\tau_\pi = 2\eta_v\beta_2$; where $\beta_0$ and $\beta_2$
are second-order transport coefficients in the entropy current. $D=u\cdot \partial$
is the time derivative in the local comoving frame and $\theta = \nabla\cdot u$ is the local
expansion rate. In the following, we shall neglect the
bulk viscosity which has negligibly small values at all temperatures, other than at the critical 
temperature $T_c = 170$ MeV for deconfinement transition. 

For the dissipative equation, we shall also consider the second-order equation for shear tensor
in the Chapman-Enskog approach obtained by iteratively solving the Boltzmann equation
in relaxation time approximation \cite{Jaiswal:2013npa,Bhalerao:2013pza,Chattopadhyay:2014lya}, 
\begin{equation}\label{SOSHEAR}
\dot{\pi}^{\langle\mu\nu\rangle} \!+ \frac{\pi^{\mu\nu}}{\tau_\pi}\!= 
2\beta_{\pi}\sigma^{\mu\nu}
\!+2\pi_\gamma^{\langle\mu}\omega^{\nu\rangle\gamma}
\!-\frac{10}{7}\pi_\gamma^{\langle\mu}\sigma^{\nu\rangle\gamma} 
\!-\frac{4}{3}\pi^{\mu\nu}\theta,
\end{equation}
where the vorticity $\omega^{\mu\nu}\equiv(\nabla^\mu u^\nu-\nabla^\nu u^\mu)/2$ 
and $\beta_\pi = 4p/5$.

The total energy-momentum (ignoring bulk viscosity) in presence of 
a noise tensor $\Xi^{\mu\nu}$ is,  
\begin{align}\label{FTt:eq}
T^{\mu\nu} = \epsilon u^\mu u^\nu - p\Delta^{\mu\nu} + \pi^{\mu\nu} + \Xi^{\mu\nu},
\end{align}
where $\Xi^{\mu\nu}(x)$ is a stochastic field in space-time with
ensemble average $\langle \Xi^{\mu\nu}(x) \rangle = 0$. The auto-correlation 
$ \langle \Xi^{\mu\nu}(x_1) \Xi^{\alpha\beta}(x_2) \rangle $ 
is derived using the fluctuation-dissipation theorem and depends on the form of 
evolution equation of the shear stress tensor. We use the theory of quasi-stationary 
fluctuations \cite{Landau} in which one considers the set of time evolution equations for the variables $x_a$,
    \begin{equation}\label{QSF:eq}
     \dot{x}_a = - \sum_a \gamma_{ab}X_b + y_a .
     \end{equation}
$X_b$ are ``driving'' forces, $y_a$ are random fluctuations.
The rate of change of entropy $S(x_a)$ is given by 
    \begin{equation}\label{sdot:eq}
     \dot{S} = - \sum_a \dot{x}_a X_a,
    \end{equation}
where $X_a = - \partial S/\partial x_a$.
As the probability of fluctuating variables in thermal equilibrium must be 
$e^S$, the autocorrelation of the noise should have the form, 
  \begin{equation}\label{FDT1:eq}
   \langle y_a(t_1)y_b(t_2) \rangle = (\gamma_{ab} + \gamma_{ba})\delta(t_1 - t_2).
  \end{equation}

The fluctuation-dissipation relation in viscous hydrodynamics can be derived 
using the above formalism \cite{Kapusta:2011gt,Chattopadhyay:2017rgh,Kumar:2013twa}.
For the MIS theory, the expression for second-order entropy four-current is given by, 
\begin{equation}\label{entrcurr:eq}
S^{\mu} = s u^{\mu} - \frac{\beta_2}{2T} u^{\mu} \pi^{\alpha\beta} \pi_{\alpha\beta},
\end{equation}
where $s = (\epsilon + p)/T$ and $\beta_2 = 1/(2\beta_\pi)$.
Using the temporal-component of $\partial_{\mu} S^{\mu}$, we obtain the rate of change of entropy,
\begin{align}\label{sdot1:eq}
\frac{dS}{dt} = \int d^3 \! x \frac{\pi^{\mu\nu}}{T} \left[ \nabla_{\mu} u_{\nu} - \beta_2
\dot{\pi}_{\mu\nu} - \beta_2 \frac{4}{3}\theta \pi_{\mu\nu}  \right].
\end{align}
In analogy to Eq. (\ref{sdot:eq}) we identify
\begin{align} \label{ide}
\dot x_a \to & \pi^{\mu\nu}, \\ 
X_a \to &  - \frac{1}{T} \left[\nabla_\mu u_\nu - \beta_2\dot{\pi}_{\mu\nu} 
- \beta_2 \lambda_\pi \theta \pi_{\mu\nu} \right] \,\, \Delta V \equiv X_{\mu\nu}. 
\end{align}
As in Eq. (\ref{QSF:eq}), we add a stochastic tensor $\xi^{\mu\nu}$ to 
the shear stress tensor, 	
\begin{align}
\pi^{\mu\nu} &= - \gamma^{\mu\nu\alpha\beta} X_{\alpha\beta} + \xi^{\mu\nu},
\end{align}	
where $\gamma^{\mu\nu\alpha\beta}$ should yield the shear
tensor expression of Eq. (\ref{IS:eq}) on contraction with $X_{\alpha\beta}$. 
Due to symmetries of $\pi^{\mu\nu}$, we have  
$\gamma^{\mu\nu\alpha\beta} = \gamma^{\nu\mu\alpha\beta}$, 
$\gamma^{\mu\alpha\beta}_{\mu} = 0$, and 
$\gamma^{\mu\nu\alpha\beta} u_\mu = 0$.
Note that the identification of $X_{\mu\nu}$ is not unique as the transformation 
$X_{\mu\nu} \to X_{\mu\nu} + H_{\mu\nu}$, 
keeps $dS/dt$ invariant if $H_{\mu\nu}$ is orthogonal to $\pi^{\mu\nu}$.
We thus have to find an autocorrelation which is insensitive to such
transformations, namely, 
$\gamma^{\mu\nu\alpha\beta} = \gamma^{\mu\nu\beta\alpha}$, 
$\gamma^{\mu\nu\alpha}_{\alpha} = 0$, and 
$\gamma^{\mu\nu\alpha\beta} u_\alpha = 0$.
 
The form of $\gamma^{\mu\nu\alpha\beta}$ consistent with the 
constraints is,
\begin{align}\label{FDTIS:eq}
\gamma^{\mu\nu\alpha\beta} =& 2 \eta_v T  \Delta^{\mu\nu\alpha\beta} .
\end{align}
Correspondingly one obtains the noise autocorrelation 
in the MIS theory to be \cite{Chattopadhyay:2017rgh}:
\begin{align}\label{autocorrIS:eq}
\langle \xi^{\mu\nu}(x) \xi^{\alpha\beta}(x') \rangle 
= & 4\eta_v T  \Delta^{\mu\nu\alpha\beta} \delta^4(x-x').
\end{align}

We now present the fluctuation-dissipation relation for Chapman-Enskog case.
The entropy four-current obtained from Boltzmann's H-theorem has the expression 
as of Eq. (\ref{entrcurr:eq}), see Ref. \cite{Chattopadhyay:2014lya}.
Following the same procedure as above, the form of $\gamma^{\mu\nu\alpha\beta}$ 
for the Chapman-Enskog case is,
\begin{align}\label{FDTCE:eq}
\gamma^{\mu\nu\alpha\beta} =& 2 \eta_v T \Big( \Delta^{\mu\nu\alpha\beta} 
- \frac{10}{7}\beta_2 \Delta^{\mu\nu}_{\zeta\kappa} \pi^\zeta_\gamma 
\Delta^{\kappa\gamma\alpha\beta} \nonumber \\ 
& + 2 \tau_\pi \Delta^{\mu\nu}_{\zeta\kappa} \omega^\zeta_\gamma 
\Delta^{\kappa\gamma\alpha\beta} \Big).
\end{align}
Consequently, one obtains the noise autocorrelation 
in the Chapman-Enskog theory to be \cite{Chattopadhyay:2017rgh}:
\begin{align}\label{autocorrCE:eq}
	\langle \xi^{\mu\nu}(x) \xi^{\alpha\beta}(x') \rangle 
= & 4\eta_v T \Big( \Delta^{\mu\nu\alpha\beta}  
- \frac{5}{7}\beta_2  \Delta^{\mu\nu}_{\zeta\kappa}\pi^\zeta_\gamma 
\Delta^{\kappa\gamma\alpha\beta} \nonumber \\ 
& - \frac{5}{7}\beta_2  \Delta^{\alpha\beta}_{\zeta\kappa}\pi^\zeta_\gamma 
\Delta^{\kappa\gamma\mu\nu} \nonumber \\
& + \omega-{\rm terms} \Big) \delta^4(x-x').
\end{align}
As opposed to the MIS case where the auto-correlation function in the fluid rest 
frame depends only on the coefficient of shear viscosity $\eta_v$ and 
temperature, the above result shows that the auto-correlation function of thermal 
noise in Chapman-Enskog scenario is sensitive 
to components of shear stress tensor as well as vorticity.  

It is important to note that in the derivation of fluctuation-dissipation relation using the 
theory of quasi-stationary fluctuations, 
the noise tensor $\xi^{\mu\nu}$ should be added to the shear evolution equation. 
However, by defining $\pi'^{\mu\nu} \equiv \pi^{\mu\nu} - \Xi^{\mu\nu}$, so as to 
obtain the same form as in Eq. (\ref{FTt:eq}), we get a relaxation-type evolution of $\Xi^{\mu\nu}$,
which essentially implies that $\Xi^{\mu\nu}$ becomes a colored-noise 
(correlated over space-times), as opposed to its uncorrelated (white-noise) structure
in the first-order Navier-Stokes theory \cite{Kapusta:2011gt}.  
In the MIS theory, we have the equation of motion of the noise tensor $\Xi^{\mu\nu}$,
\begin{align}
\dot \Xi^{\langle\mu\nu\rangle} = - \frac{1}{\tau_\pi} 
\left( \Xi^{\mu\nu} - \xi^{\mu\nu} \right) - \frac{4}{3} \Xi^{\mu\nu}\theta,  
\end{align}
and for the CE equation we get
\begin{align}
\dot \Xi^{\langle\mu\nu\rangle} = - \frac{1}{\tau_\pi} 
\left( \Xi^{\mu\nu} - \xi^{\mu\nu} \right) + 2\pi_\gamma^{\langle\mu}\omega^{\nu\rangle\gamma}
- \frac{10}{7} \Xi^{\langle\mu}_\gamma \sigma^{\nu \rangle\gamma}
- \frac{4}{3}\Xi^{\mu\nu}\theta. 
\end{align}

In the following, we shall linearize the stochastic hydrodynamic equations about a background (averaged)
solution such that the r.h.s of Eqs. (\ref{autocorrIS:eq}) and (\ref{autocorrCE:eq}) will be computed using these averaged solutions.
Considering fluctuations in temperature (or energy density), flow velocity and shear pressure tensor 
\cite{Kapusta:2011gt}, their values in the linearized limit can be written as  
\begin{align}\label{nonlin:eq}
\epsilon &=  \epsilon_0 + \delta\epsilon \equiv  \epsilon_0 + \epsilon_1, \nonumber\\
u^\mu &=  u_0^\mu + \delta u^\mu  \equiv   u_0^\mu + u_1^\mu, \nonumber\\
\pi^{\mu\nu} &=  \pi_0^{\mu\nu} + \delta\pi^{\mu\nu}  \equiv  \pi_0^{\mu\nu} +  \pi_1^{\mu\nu}.  
\end{align}
The subscript ``0" corresponds to the average (noiseless) values of the quantities whose evolution will be
described in Sec. II B. The equations for fluctuations (denoted by subscript ``1") are presented in Sec. II C. 
As a consequence of Eq. (\ref{nonlin:eq}), the total energy-momentum tensor 
of Eq.  (\ref{FTt:eq}) can be decomposed into 
$T^{\mu\nu} = T_0^{\mu\nu} + T_1^{\mu\nu}$, consisting of a noiseless part $T_0^{\mu\nu}$ 
and a fluctuating part $\delta T^{\mu\nu} \equiv T_1^{\mu\nu}$.

\subsection{(1+1)D non-boost-invariant viscous hydrodynamic}

In the present calculation for non-boost-invariant longitudinal expansion, the hydrodynamic
equations effectively correspond to (1+1)D. For high-energy collisions at RHIC and LHC, the
space-time evolution can be conveniently described in the Milne coordinates of longitudinal
proper time $\tau = \sqrt{t^2-z^2}$ and space-time rapidity $\eta = \ln[(t+z)/(t-z)]/2$.
In ($\tau,x,y,\eta$)
coordinates, the metric tensor becomes $g^{mn} = {\rm diag}(1,-1,-1,-1/\tau^2)$. Due to the
translational and rotational invariance in the transverse plane, the four-velocity can be
parametrized as $u_0^\mu = (u_0^\tau, u_0^x, u_0^y, u_0^\eta) \equiv \gamma_0 (1,0,0,v_{\eta_0})$,
where $\gamma_0 = 1/\sqrt{1 - \tau^2 v_{\eta_0}^2}$ arises from the normalization condition
$u_0^\mu u_{0\mu}=1$. The nonvanishing components of the Christoffel symbols are 
$\Gamma^\eta_{\eta\tau} = \Gamma^\eta_{\tau\eta} = 1/\tau$ and $\Gamma^\tau_{\eta\eta} = \tau$.
The time derivative in the local fluid rest frame and the local expansion rate are then
\begin{align}\label{Dtheta}
D_0 &= u_0 \cdot \partial = \gamma_0 \left(\partial_\tau + v_{\eta_0} \partial_\eta \right), \\
\theta_0 &= \partial \cdot u_0 = \frac{1}{\tau} \partial_\tau(\tau\gamma_0) + \partial_\eta(\gamma_0 v_{\eta_0}) .
\end{align}

Due to constraints on $\pi_0^{\mu\nu}$, namely, orthogonality to $u_0^{\mu}$ (i.e. $\pi_0^{mn}u_{0_m} = 0$), 
tracelessness ($\pi_0^{mn}g_{mn} = 0$), and azimuthal ($x-y$) symmetry, 
there is only one independent component which we take to be $\pi_0^{\eta\eta}$.
The other non-vanishing components can be expressed in terms of $\pi_0^{\eta\eta}$ as,
\begin{align}\label{ortho:eq}
\pi_0^{\tau\tau} &= \tau^4 v_{\eta_0}^2 \pi_0^{\eta\eta}, ~~
\pi_0^{\tau\eta} = \tau^2 v_{\eta_0} \pi_0^{\eta\eta}, ~~ \nonumber\\ 
\pi_0^{xx}&=\pi_{0}^{yy} = -\frac{\tau^2 }{2 \gamma_0^2}\pi_{0}^{\eta\eta}.
\end{align}

The two independent components of the noiseless part of energy-momentum tensor $T_0^{\mu\nu}$ 
in the global frame reduce to
\begin{align}\label{Ttt:eq}
T_0^{\tau\tau} &= (\epsilon_0 + p_0)\gamma_0^2 - p_0 + \pi_0^{\tau\tau} 
= (\epsilon_0 + {\cal P}_0)\gamma_0^2 - {\cal P}_0, \\ 
T_0^{\tau\eta} &= (\epsilon_0 + p_0)\gamma_0^2 v_{\eta_0} + \pi_0^{\tau\eta} 
= (\epsilon_0 + {\cal P}_0)\gamma_0^2 v_{\eta_0}, \label{Ttn:eq}
\end{align}
where Eq. (\ref{ortho:eq}) has been used to obtain the right-hand-side of the second
equality. The effective (longitudinal) pressure is denoted as 
${\cal P}_0 = p_0 + \tau^2 \pi_0^{\eta\eta}/\gamma_0^2$.
Using $T_0^{\eta\eta} = [(\epsilon_0 + {\cal P}_0)\gamma_0^2 - \epsilon_0]/\tau^2$
the equation of motion $T^{mn}_{;m}=0$ for the $n=\tau$ and $n=\eta$ component can be
written as 
\begin{align}\label{Tevol1:eq}
& \partial_\tau(\tilde T_0^{\tau\tau}) + \partial_\eta (\tilde v_{\eta_0} \tilde T_0^{\tau\tau}) 
= - \left(\epsilon_0 + {\cal P}_0\right)\gamma_0^2 + \epsilon_0 , \\
& \partial_\tau(\tilde T_0^{\tau\eta}) + \partial_\eta (v_{\eta_0} \tilde T_0^{\tau\eta}  + 
\tilde {\cal P}_0/\tau^2)  = - 2\left(\epsilon_0 + {\cal P}_0\right)\gamma_0^2 v_{\eta_0} . \label{Tevol2:eq}
\end{align}
Here we have used the shorthand notation $\tilde A^{mn} = \tau A^{mn}$ and 
$\tilde v_{\eta_0} = \tilde T_0^{\tau\eta}/ \tilde T_0^{\tau\tau} =  T_0^{\tau\eta}/T_0^{\tau\tau}$.  

In order to write the evolution equation for $\pi_0^{\eta\eta}$, we obtain the general 
form of relaxation equation for the full tensor $\dot{\pi}^{\langle \eta \eta \rangle}$.
Using the orthogonality conditions $\pi^{\mu\nu} u_{\nu} = 0$ and $\dot{u}^{\mu} u_{\mu} = 0$, 
the comoving derivative of $\pi^{\eta\eta}$ in (1+1)D can be expressed in a compact form,
\begin{align}\label{pieta:Eq}
\dot{\pi}^{\langle \eta\eta \rangle} = \frac{\gamma^2}{\tau^2} u^{\mu} 
\partial_{\mu} \left( \frac{\tau^2 \pi^{\eta\eta}}{\gamma^2} \right) .
\end{align}
Moreover, using $\sigma^{\eta\eta} = - (2/3)(\gamma^2 \theta /\tau^2)$
and defining the total (background plus noise) shear stress $\pi = - \tau^2 \pi^{\eta\eta}/\gamma^2$, 
the MIS Eq. (\ref{IS:eq}) on adding the noise term $\xi^{\mu\nu}$ can be written as
\begin{align}\label{pievol:eq}
 u^\mu \partial_\mu \pi 
= -\frac{1}{\tau_\pi} \left( \pi + \frac{\tau^2}{\gamma^2} \xi^{\eta\eta} 
- \frac{4}{3} \eta_v \: \theta \right) - \lambda_\pi \pi \: \theta.
\end{align}
Here the coefficient $\lambda_\pi = 4/3$ in this MIS dissipative equation.
The evolution equation for the background shear stress component 
$\pi_0 \equiv - \tau^2\pi_0^{\eta\eta}/\gamma_0^2$ then has the form, 
\begin{align}\label{pi_0:Eq}
u_0^\mu \partial_\mu \pi_0 
= -\frac{1}{\tau_\pi} \left( \pi_0 - \frac{4}{3} \eta_v \: \theta_0 \right)
- \lambda_\pi \pi_0 \: \theta_0.
\end{align}
For the CE case the dissipative equation (\ref{SOSHEAR}) in (1+1)D has the same 
form as Eq. (\ref{pievol:eq}) with $\lambda_\pi = 38/21$.

The three evolution Eqs. (\ref{Tevol1:eq}), (\ref{Tevol2:eq}), (\ref{pi_0:Eq}) in four unknowns are closed
with the equation of state $p = p(\epsilon)$. 
Using Eqs. (\ref{Ttt:eq})-(\ref{Ttn:eq}),
one can express the energy density and the (longitudinal) velocity as
\begin{align}\label{envel0:eq}
\epsilon_0 &=  T_0^{\tau\tau} - \tau^2 v_{\eta_0} T_0^{\tau\eta}, \\
v_{\eta_0} &= \frac{T_0^{\tau\eta}}{T_0^{\tau\tau} + 
p_0\left(\epsilon_0 = T_0^{\tau\tau} - \tau^2 v_{\eta_0} T_0^{\tau\eta}\right) - \pi_0 }, 
\end{align}
and these allow one to extract $v_{\eta_0}$ by one-dimensional zero-search. The above set of evolution
equations are solved using SHASTA-FCT algorithm.

\subsection{Thermal fluctuations in non-boost-invariant viscous hydrodynamics}

We will now obtain the linearized hydrodynamic equations for thermal fluctuations in the non-boost-invariant 
(1+1)D expansion of matter within MIS formulation. Substituting the first-order fluctuations 
of Eq. (\ref{nonlin:eq}) into the total energy-momentum tensor: 
\begin{align}\label{FT:eq}
T^{\mu\nu} &= \epsilon u^\mu u^\nu - p\Delta^{\mu\nu} + \pi^{\mu\nu}, \nonumber\\  
&= T_0^{\mu\nu} + \delta T_{\rm id}^{\mu\nu} + \delta\pi^{\mu\nu}
= T_0^{\mu\nu} + T_1^{\mu\nu}, 
\end{align}
where $T_0^{\mu\nu}$ is the noiseless energy-momentum tensor 
whose evolution equations has been obtained in Sec. II B.
Note that the noise term $\xi^{\mu\nu}$ has been included in the shear
evolution Eq. (\ref{pievol:eq}).
The fluctuating part of the ideal energy-momentum tensor is 
$\delta T_{\rm id}^{\mu\nu} = \delta(\epsilon u^\mu u^\nu - p\Delta^{\mu\nu})$, and can be 
determined by the fluctuating variables ($\delta\epsilon, \delta u^\mu, \delta p$). 
The conservation equations for the total energy-momentum tensor, $\partial_\mu T^{\mu\nu}=0$,
along with that for the average part, $\partial_\mu T_0^{\mu\nu}=0$, lead to
\begin{align}\label{FCT:eq}
\partial_\mu (\delta T_{\rm id}^{\mu\nu} + \delta\pi^{\mu\nu})
\equiv \partial_\mu (\delta T^{\mu\nu}) = 0.
\end{align}
The above equations are combined to yield
\begin{align}\label{FTderiv:eq}
\delta T^{\mu\nu} \equiv T_1^{\mu\nu} =& 
\epsilon_1 u_0^\mu u_0^\nu + \epsilon_0 u_1^\mu u_0^\nu + \epsilon_0 u_0^\mu u_1^\nu \nonumber\\
& - p_0\Delta_1^{\mu\nu} - p_1\Delta_0^{\mu\nu} + \pi_1^{\mu\nu}.
\end{align}
While the event-averaged fluctuations yield $\langle \delta T^{\mu\nu} \rangle = 0$, 
thermal noise $\xi^{\mu\nu}$ 
induces a nonvanishing $\langle \delta T^{\mu\nu} \delta T^{\alpha\beta} \rangle$
which was demonstrated to result in the two-particle rapidity correlation 
\cite{Kapusta:2011gt} and affect the event-by-event fluctuation of elliptic 
flow \cite{Young:2014pka}.

On imposing the orthonormality condition of the total four-velocity of the fluid, i.e. 
$u^\mu u_\mu = (u_0^\mu + u_1^\mu)(u_{0\mu} + u_{1\mu}) = 1$ and noting that
$u_0^\mu u_{0\mu} =1$, we get in the linearized limit 
\begin{align}\label{Fu1:eq}
u_1^\tau = \tau^2 u_1^\eta v_{\eta_0}.
\end{align}
Further, by using the orthogonality of the total shear pressure with  
four-velocity, $\pi^{\mu\nu} u_\nu =0$,
and the corresponding Eq. (\ref{ortho:eq}) for the noiseless part, one gets
for the fluctuating shear stress component
\begin{align}\label{Fortho:eq}
\pi_1^{\mu\nu} u_{0\nu}  = - \pi_0^{\mu\nu} u_{1\nu}.
\end{align}
Making use of these conditions, together with the tracelessness of $\pi^{\mu\nu}$ and 
Eqs. (\ref{ortho:eq}), we obtain for the (1+1)D expansion,
\begin{align}\label{Fpi:eq}
\pi_1^{\tau\tau} &=  \tau^4 v_{\eta_0}^2 \pi_1^{\eta\eta}  
- 2 \frac{\tau^2 v_{\eta_0} u^{\eta}_1}{\gamma_0} \pi_0 , \nonumber\\
\pi_1^{\tau\eta} &= \tau^2 \pi_1^{\eta\eta} v_{\eta_0} - \frac{u^{\eta}_1}{\gamma_0} \pi_0.
\end{align}

The corresponding fluctuating components of the energy-momentum tensor of Eq. (\ref{FTderiv:eq}) is then:
\begin{align}\label{FTmunu:eq}
T_1^{\tau\tau} &= (\epsilon_1 + \tau^2 {\cal P}_1)\gamma_0^2 - \tau^2 {\cal P}_1 
+\tau^2 u_1^\eta v_{\eta_0} \gamma_0 {\cal U}_0 (2 - \tau^2 v_{\eta_0}^2), \nonumber \\
T_1^{\tau\eta} &= (\epsilon_1 + \tau^2 {\cal P}_1)\gamma_0^2 v_{\eta_0} 
+ u_1^\eta \gamma_0 {\cal U}_0, \nonumber \\
T_1^{\eta\eta} &= T_1^{\tau\eta} v_{\eta_0}  + {\cal P}_1.
\end{align}
We use the definition ${\cal P}_1 = u_1^\eta v_{\eta_0} {\cal U}_0 /\gamma_0 + {\cal V}_1/\tau^2$ with
${\cal U}_0 = \epsilon_0 + p_0 - \pi_0$ and ${\cal V}_1 = p_1 - \pi_1$. Here 
$\pi_1 \equiv -(\tau^2/\gamma_0^2) \pi_1^{\eta\eta} - 2 (\tau^{2}u^\eta_1/\gamma_0) v_{\eta_0} \pi_0$
is obtained by linearizing the total shear tensor 
$\pi = -\tau^2 \pi^{\eta\eta}/\gamma^2$ defined in Sec. II B.
The equations for the noise part of the energy-momentum conservations are then:
\begin{align}\label{FTevol:eq}
\partial_\tau(\tilde T_1^{\tau\tau}) + \partial_\eta (\tilde v_{\eta_1} \tilde T_1^{\tau\tau}) 
=& \epsilon_1 -  \left(\epsilon_1 + \tau^2 {\cal P}_1\right)\gamma_0^2 \nonumber\\
& - \tau^2 u_1^\eta v_{\eta_0} \gamma_0 {\cal U}_0, \\
\partial_\tau(\tilde T_1^{\tau\eta}) + \partial_\eta (v_{\eta_0} \tilde T_1^{\tau\eta}  
+ \tilde {\cal P}_1) 
=& - 2\left(\epsilon_1 + \tau^2 {\cal P}_1\right)\gamma_0^2 v_{\eta_0} \nonumber\\
& - 2 u_1^\eta \gamma_0 {\cal U}_0.
\end{align}
The stochastic MIS equations for the noise term in the linearized limit can be
obtained from Eq. (\ref{IS:eq}).
The dissipative equation for the independent component $\pi_1$ then reads,
\begin{align}\label{Fpievol:eq}
\gamma_0 \left( \partial_\tau + v_{\eta_0}\partial_\eta \right) \pi_1 
=& \frac{1}{\tau_\pi}\left[ - \pi_1 + \xi
 + \frac{4\eta_v}{3s} \left( s_0 \theta_1 + s_1 \theta_0 \right) \right] \nonumber\\
&  - u_1^\eta \left( \tau^2 v_{\eta_0} \partial_\tau \pi_0 + \partial_\eta \pi_0 \right) \nonumber\\
& - \lambda_\pi \left( \theta_0 \pi_1 + \theta_1 \pi_0 \right),
\end{align}
where the local expansion rate for the velocity fluctuation is of the form
$\theta_1 = (1/\tau) \partial_\tau(\tau u_1^\tau) + \partial_\eta u_1^\eta$
and we have defined $\xi = - \tau^2 \xi^{\eta\eta}/\gamma_0^2$.
The autocorrelation for the noise term in general is found to be 
\begin{align}\label{noiseIS:eq}
\langle \xi(\tau_1,\eta_1) \xi(\tau_2,\eta_2) \rangle =& \frac{8\eta_v T}{3 \tau A_\perp} \Big[ 1 - {\cal A}\beta_2 \pi_0 \Big] \nonumber \\
 & \times \delta(\tau_1 - \tau_2) \delta(\eta_1 - \eta_2),
\end{align}
where ${\cal A} = 0 $ in MIS case and
${\cal A} = 5/7$ in CE formalism, and the delta function in the transverse direction
$\delta({\bf x} - {\bf x'})\delta({\bf y} - {\bf y'}) = 1/A_\perp$ is represented by the inverse
of the effective (overlap) transverse area $A_\perp$ of the colliding nuclei.

By imposing the Landau-matching condition for the total energy-momentum tensor, 
$T^{\mu\nu}u_\nu = \epsilon u_\mu$,
and also for the average part, one can determine the fluctuating energy density and velocity as:
\begin{align}\label{envel1:eq}
\epsilon_1 &=  \left( T_1^{\tau\tau} - \tau^2 v_{\eta_0} T_1^{\tau\eta} \right) 
+ \frac{\tau^2 u_1^\eta}{\gamma_0} 
\left( v_{\eta_0} T_0^{\tau\tau} - T_0^{\tau\eta} - \epsilon_0 v_{\eta_0} \right), \\
u_1^\eta &= \frac{\gamma_0}{{\cal U}_0} \left( 
T_1^{\tau\eta} - v_{\eta_0} T_1^{\tau\tau} - v_{\eta_0} {\cal V}_1 \right).
\end{align}
which can be obtained from one-dimensional root search method.
The hydrodynamic fluctuation Eqs. (\ref{FTevol:eq})-(\ref{Fpievol:eq})
are solved perturbatively in $\tau-\eta$ coordinates using the MacCormack (predictor-corrector) 
method.

\subsection{Freeze-out and two-particle rapidity correlations}

We shall now consider the freeze-out of a fluid system that undergoes (nonequilibrium) 
viscous evolution with thermal fluctuations. The freeze-out of a near-thermalized
fluid to a free-streaming (noninteracting) particles is obtained via the standard
Cooper-Frye prescription \cite{Cooper:1974mv}. We will consider isothermal freeze-out 
that corresponds to a freeze-out from a hypersurface $\Sigma(x)$ when its temperature drops below a
critical (decoupling) value of $T_{\rm dec}$. The particle spectrum can be 
obtained as 
\begin{align}\label{CF:eq}
E\frac{dN}{d^3p} = \frac{g}{(2\pi)^3} \int_\Sigma  d\Sigma_\mu p^\mu f(x,p),
\end{align}
where $p^\mu$ is the four-momentum of the particle with degeneracy $g$, 
$d\Sigma^\mu$ is the outward-directed normal vector on an infinitesimal element 
of the hypersurface $\Sigma(x)$.

In the present ($\tau,x,y,\eta$) coordinate system, the three-dimensional
volume element at freeze-out is  
\begin{align}\label{volel}
d\Sigma_\mu \equiv& [d\Sigma_\tau(\eta), d\Sigma_x(\eta), d\Sigma_y, d\Sigma_\eta(\eta)],  \nonumber\\
=&  \Big(1, 0,0, -\frac{\partial \tau_f}{\partial \eta} \Big) \tau_f d\eta d{\bf x}_\perp.
\end{align}
where $\tau_f(\eta)$ is the freeze-out time at the decoupling temperature $T_{\rm dec}$.
The particle four-momentum,
$p^\mu \equiv (p^0, p^x, p^y, p^z) = (m_T\cosh y, p_x, p_y, m_T \sinh y)$, 
in ($\tau,x,y,\eta$) coordinates becomes 
\begin{align}\label{4mom}
p^\mu = [m_T\cosh(y-\eta), {\bf p}_\perp,  m_T\sinh(y-\eta)].
\end{align}
Here $m_T = \sqrt{p_T^2 + m^2}$ is the transverse mass of the particle with transverse 
momentum $p_T$ and kinematic rapidity $y =\tanh^{-1}(p^z/p^0)$.  The integration measure 
at the constant temperature freeze-out hypersurface $\Sigma(x)$ is then
$p^\mu d\Sigma_\mu = d\eta d{\bf x}_\perp m_T 
\: \partial[-\tau_f \sinh(y-\eta)]/\partial\eta$.

The phase-space distribution function at freeze-out, 
$f(x,p) = f_{\rm eq}(x,p) + f_{\rm vis}(x,p)$ consists of equilibrium contribution 
\begin{align}\label{feq:eq}
f_{\rm eq} = {\rm exp}[p \cdot u/T \pm 1]^{-1} \approx {\rm exp}(-p\cdot u)/T, 
\end{align}
and the nonequilibrium viscous correction, which has the form
derived from the Grad's 14-moment approximation \cite{Grad}:
\begin{align}\label{fvis:eq}
f_{\rm vis} = f_{\rm eq}(1 \mp f_{\rm eq}) \frac{p^\mu p^\nu \pi_{\mu\nu}}{2(\epsilon+p)T^2} 
\approx  f_{\rm eq} \frac{p^\mu p^\nu \pi_{\mu\nu}}{2(\epsilon+p)T^2}. 
\end{align}
Note that the total flow 
velocity $u^\mu \equiv u^\mu(\tau_f,\eta)$ and the total temperature 
$T \equiv T(\tau_f,\eta)$ are evaluated at the freeze-out hypersurface coordinates.  

In order to evaluate Eq. (\ref{CF:eq}), we note 
that the total distribution function $f(x,p)$ 
has contributions from the average (noiseless) and the thermal noise 
parts. In the linearized limit, $f$ can be written as
\begin{align}\label{f:eq}
f(x,p) = f_0(x,p) + \delta f(x,p) \equiv  f_0(x,p) + f_1(x,p).
\end{align}
As a consequence of Eq. (\ref{nonlin:eq}), one can write the average part
of the distribution function as \cite{Chattopadhyay:2017rgh}
\begin{align}\label{f0:eq}
f_0 =& (f_{\rm eq})_0 
\left( 1 + {K_0}_{\mu \nu} \pi_0^{\mu\nu} \right),
\end{align}
where $K_0^{\mu\nu} = p^\mu p^\nu[2(\epsilon_0+p_0)T_0^2)]^{-1}$, and  
the total temperature as $T=T_0 + \delta T \equiv T_0 + T_1$. 
The noise part $f_1$ has contribution from ideal as well as viscous fluctuations
\begin{align}\label{f1:eq}
f_1 =& (f_{\rm eq})_1  + {K_0}_{\mu\nu} \Big[ 
(f_{\rm eq})_1 \pi_0^{\mu\nu} + (f_{\rm eq})_0 \pi_1^{\mu\nu} \nonumber \\
& + (f_{\rm eq})_0 \pi_0^{\mu\nu}
\left( \frac{T_1}{T_0} + \frac{\epsilon_1+p_1}{\epsilon_0+p_0} \right)  \Big] ,
\end{align}
where $(f_{\rm eq})_0 = {\rm exp}(-u_0^\mu p_\mu/T_0)$ 
and $(f_{\rm eq})_1 = (f_{\rm eq})_0 (T_1 u_0^\mu p_\mu/T_0^2 - u_1^\mu p_\mu/T_0)$
are, respectively, the noiseless and the noise parts of the equilibrium (ideal) 
distribution function. The terms within the square brackets 
in Eq. (\ref{f1:eq}) refer to contributions from viscous fluctuations.

The rapidity distribution of the particle, corresponding to Eq. (\ref{CF:eq}), 
then reduces to 
\begin{align}\label{rapidity:eq}
\frac{dN}{dy} =& \frac{g T_0^3 A_\perp}{(2\pi)^3} \int d\eta \: S(y,\eta) \nonumber \\ 
&\times \int dp_x dp_y \: m_T [f_0(x,p) + \delta f(x,p)] \nonumber\\
& \equiv (dN/dy)_0  + \delta (dN/dy) .
\end{align}
Here $A_\perp = \int d{\bf x}_\perp$ is the usual transverse area of Eq. (\ref{noiseIS:eq}) and 
$S(y,\eta) \equiv \partial[-\tau_f \sinh(y-\eta)]/\partial\eta$.
For the non-boost-invariant longitudinal flow, the averaged particle rapidity distribution
corresponding to Eq. (\ref{CF:eq}) becomes
\begin{align}\label{frap:eq}
\left(\frac{dN}{dy}\right)_{\!\! 0}  =& ~\frac{g T_0^3 A_\perp}{(2\pi)^2} 
\int \frac{d\eta} {\cosh^3 \Lambda}  S(y,\eta)  
\Big[ \Gamma_3(\Lambda) \nonumber \nonumber \\
& + \frac{\pi_0}{4 w_0} 
\left({\cal C}(\Lambda) \Gamma_5(\Lambda) 
 - \frac{m^2}{T_0^2}\Gamma_3(\Lambda) \right) \Big].
\end{align}
We use the definition $\Lambda \equiv (y - \eta - \kappa)$, where $\kappa \equiv \tanh^{-1}{(\tau v_{\eta_0})}$
and $w_0 = (\epsilon_0 + p_0)$ is the background enthalpy density.
$\Gamma_k(\Lambda) \equiv \Gamma(k, m\cosh \Lambda /T_0)$ denotes the incomplete Gamma function
of the $k$th kind \cite{Stegun} and ${\cal C}(\Lambda) = 3~{\rm sech}^2\Lambda -2$.
Note that the second term within the square brackets stems from viscous corrections.
For the fluctuating part we have
\begin{align}
\delta\frac{dN}{dy} =& \frac{g T_0^3 A_\perp}{(2\pi)^2} 
\int  d\eta \, S(y,\eta) \: \Big[ {\cal F}_T (y,\eta) \frac{T_1(\eta)}{T_0}  \nonumber\\
&+ {\cal F}_u (y,\eta) \, \frac{\tau_f u_1^\eta(\eta)}{\gamma_0} + {\cal F}_\pi (y,\eta) \, \frac{\pi_1(\eta)}{w_0} \Big],
\end{align}
Here ${\cal F}_{T,u,\pi}$ are the coefficients of the fluctuations,
($T_1/T_0, \tau_f u_1^\eta/\gamma_0, \pi_1/w_0$), that are obtained by performing the momentum integrals:
\begin{align}\label{Coef_Tup:eq}
{\cal F}_T \cosh^3 \Lambda =&  \Gamma_4 
- \frac{\pi_0}{4w_0} \Big[ \frac{m^2}{T_0^2}\left(\Gamma_4 - \kappa \Gamma_3 \right)  \nonumber \\
-& {\cal C}(\Lambda) \left(\Gamma_6 - \kappa \Gamma_5 \right) \Big], \\
{\cal F}_u \cosh^3 \Lambda =&  \Gamma_4 \tanh\Lambda 
- \frac{\pi_0}{4w_0} \tanh\Lambda \nonumber \\
&\times \Big[ \frac{m^2}{T_0^2}\Gamma_4 - {\cal C}(\Lambda) \left( \Gamma_6  
- 2\frac{\tanh \kappa}{\tanh \Lambda} \right)     \nonumber \\ 
& +  4 \Gamma_5  \frac{2\sinh\kappa - \sinh(2\Lambda + \kappa)}{\cosh\kappa \: \sinh 2\Lambda} \Big],  \\
{\cal F}_\pi \cosh^3 \Lambda =& \frac{1}{4} \Big[ 
 {\cal C}(\Lambda) \Gamma_5 - \frac{m^2}{T_0^2}\Gamma_3 \Big],
\end{align}
where $\kappa = 2 + (T_0/w_0)\partial w_0 /\partial T_0$.

The two-particle rapidity correlator due to fluctuations can then
be written as 
\begin{align}\label{FRapCor:eq}
\left\langle \delta\frac{dN}{dy_1} \ \delta\frac{dN}{dy_2} \right\rangle =&  
\left[ \frac{g T_0^3 A_\perp}{(2\pi)^2} \right]^2  
\int d\eta_1 \, S(y_1,\eta_1) \int d\eta_2 \, S(y_2,\eta_2) \nonumber \\ 
\times & \sum_{X,Y} {\cal F}_X (y_1,\eta_1) {\cal F}_Y (y_2,\eta_2) \: \langle X(\eta_1) Y(\eta_2) \rangle .
\end{align}
Here $(X,Y) \equiv (T_1, u_1^\eta, \pi_1)$ and $\langle X(\eta_1) Y(\eta_2) \rangle$ are the two-point
correlators between the fluctuating variables calculated at the freeze-out hypersurface.
The CE formalism gives the same above expression for the two-particle correlations but 
with different coefficients ${\cal F}_X$ due to modified form of the 
viscous correction $f_{\rm vis}$ \cite{Chattopadhyay:2017rgh}.

\section{Results and discussions}

\subsection{Numerical test of the (1+1)D non-boost-invariant code}

We have developed a numerical simulation code for the non-boost-invariance longitudinal expansion 
of matter by employing the relativistic hydrodynamic equations formulated in the Milne coordinates. 
The SHASTA-FCT algorithm was used to solve the coupled conservative equations, which is
an efficient hydrodynamic shock capturing scheme. In this section we shall discuss some
numerical test of our code. In particular,  the numerical results will be compared with
the analytical solutions for one-dimensional expansion of matter, namely, the Riemann
simple wave solutions and the Landau-Khalatnikov solution
\cite{LandauS,Khalatnikov,Wong:2014sda}.

The relativistic Riemann problem \cite{Wong:2014sda,Bouras:2010hm,Okamoto:2016pbc}
can be explored by considering a hydrodynamic state 
${\cal H}(\epsilon, v^x, v^y, v^z)$, that is a function of hydrodynamic variables,  
and the state has a discontinuity at the initial time $t=t_0$ and at the location
$z=z_i$. The initial boundary-value problem can be represented 
in the Cartesian coordinate as
${\cal H}(t_0, x, y, z) \equiv {\cal H}_L$ for $z <z_i$ and  
${\cal H}(t_0, x, y, z) \equiv {\cal H}_R$ for $z >z_i$. The time evolution of the
initial disturbance originating at $z=z_i$ can be described by a Riemann simple
wave solution for one-dimension relativistic hydrodynamic expansion. In fact, the
solution corresponds to superposition of three nonlinear wave propagation.
Two of them are shock and/or rarefaction waves that are formed near the boundary $z=z_i$,
and travelling in the opposite directions with the speed of sound
$c_s = \sqrt{\partial \epsilon /\partial p}$. The other is the hydrodynamic
propagation of the discontinuity itself. In Milne coordinates, the Riemann initial-value
problem for the hydrodynamic states remain unchanged \cite{Okamoto:2016pbc}, viz.  
${\cal H}(\tau_0, x, y, \eta) = {\cal H}_L(\eta < \eta_i)$ and  
${\cal H}(\tau_0, x, y, \eta) = {\cal H}_R(\eta > \eta_i)$, and the discontinuity is now
at the space-time rapidity $\eta = \eta_i$ at the initial proper time $\tau = \tau_i$.

In contrast, the Landau-Khalatnikov solution is applicable at much later times for the
rarefaction wave propagation inside the medium. For instance, if the stopped matter
in nucleus-nucleus collisions is represented by a slab of width $z = 2\Delta$ 
(in Cartesian coordinate) in contact with vacuum on either side, then the complete 
hydrodynamical wave would be given by Riemann solution for expansion of matter 
into the vacuum at time $t \Delta/c_s$ and Landau-Khalatnikov solution for wave
inside the slab at later time $t > \Delta/c_s$.  

To test the stability of our numerical solution obtained in the Milne coordinate, we
note that the velocity  fields in the Milne and Cartesian coordinates are related by
\begin{align}\label{veta:eq}
v^\eta = \frac{1}{\tau} \frac{-\sinh\eta + v^z \cosh\eta}{\cosh\eta - v^z \sinh\eta}.
\end{align}
Thus the velocity fields in the Cartesian coordinate $u^i/u^t = (v^x,v^y,v^z)$
is independent of rapidity, while $u^m/u^\tau = (v^x,v^y,v^\eta)$ depends on rapidity.  
As a first test of our one-dimensional hydrodynamic expansion simulation, we consider
a slab situated at $|\eta| \leq 1.5$ at the initial time $\tau =\tau_0 =1$ fm/c, 
and has an energy density of $\epsilon=\epsilon_0=120$ GeV/fm$^3$. 
Initially the slab is at rest in the Cartesian coordinate ($v^z =0$) and in contact with
vacuum on both the ends. Thus the initial condition can be recast into 
\begin{align}\label{slabr:eq}
\epsilon &= 120 \: {\rm GeV/fm}^3, ~~ v^\eta = -\tanh\eta
~~~~ {\rm for} ~~ |\eta| \leq 1.5 \nonumber\\
\epsilon &= 0, ~~~~~~~~~~~~~~~~~~ v^\eta = 0 ~~~~~~~~~~~~~ {\rm for} ~~ |\eta| > 1.5 
\end{align}
With the conformal equation of state $\epsilon = 3p$ used here, the corresponding
initial temperature of the slab is $T_0 \sim \epsilon^{1/4} \approx 507$ MeV.
With these initial values, we perform hydrodynamic simulation for the time evolution,
and compare the the numerical results with the analytic Riemann solution for energy density 
\cite{Okamoto:2016pbc},
\begin{align}\label{riemann:eq}
\epsilon = \epsilon_0 \left[ \left(\frac{1+c_s}{1-c_s}\right)  
\left\{ \frac{1 + (z-z_i)/(t-t_i)}{1 - (z-z_i)/(t-t_i)} \right\} \right]^{2c_s} ,
\end{align}
where the transformations from the Cartesian to Milne coordinate are
$z = \tau \: \sinh\eta$, $t = \tau \: \cosh\eta$, and accordingly for the initial coordinates.
The numerical velocity $v^\eta$ can be also compared to the analytic solution of Eq. (\ref{veta:eq}).

\begin{figure}[t]
\includegraphics[width=\linewidth]{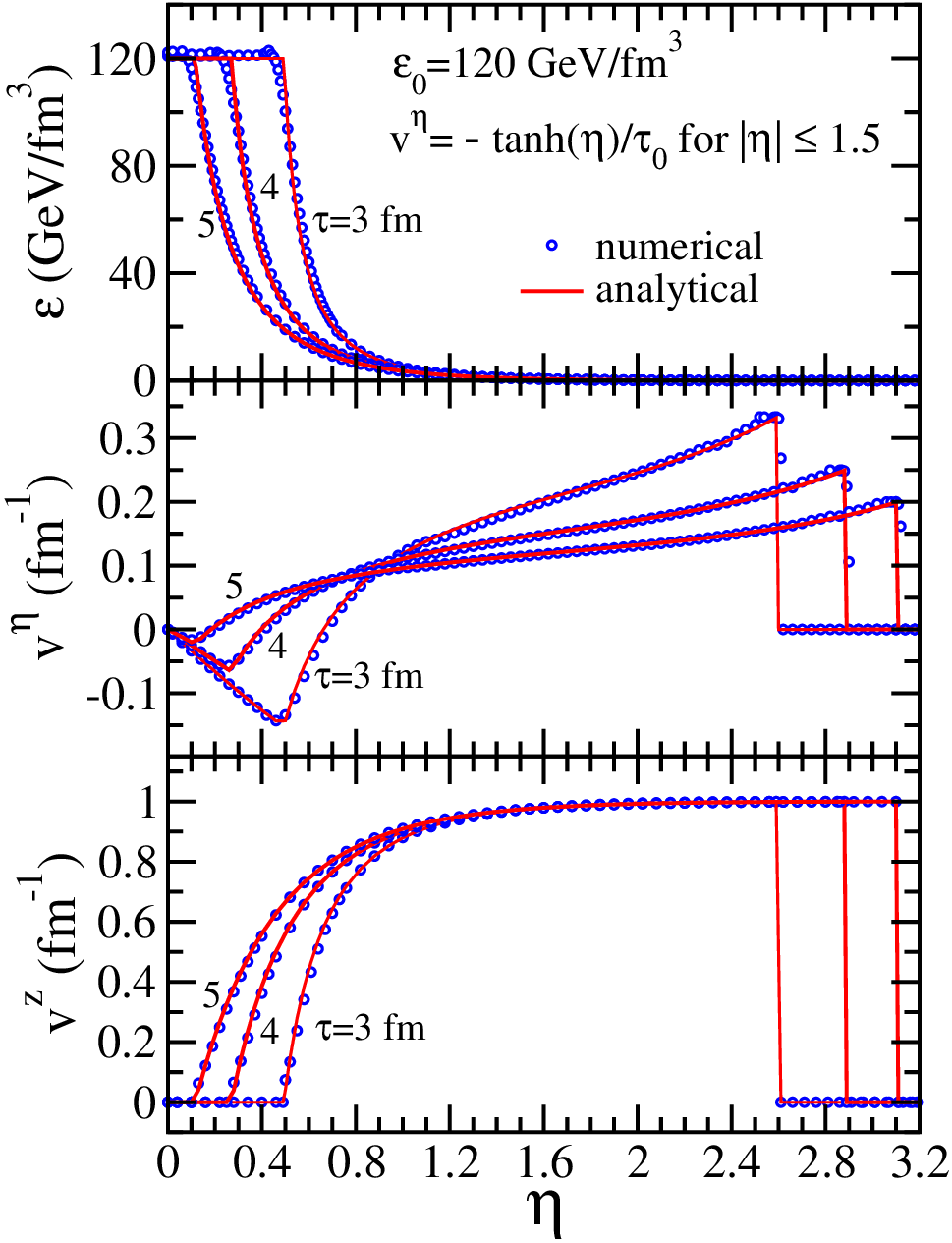}
\caption{Comparison between the Riemann analytical solutions (lines) and the numerical results
(circles) for the rapidity dependence of energy density $\epsilon$, velocities $v^\eta$
and $v^z$ at proper times $\tau = 3,4,5$ fm/c for the initial condition of Eq. (\ref{slabr:eq})
at time $\tau_0 = 1$ fm/c.}
\label{fig:Riemann}
\end{figure}

Figure \ref{fig:Riemann} shows comparison of numerical and analytic results, for the rapidity dependence
of the energy density, velocity $v^\eta$, and the velocity $v^z$ obtained from Eq. (\ref{veta:eq})
by using the corresponding $v^\eta$ values. 
All the results are at later times of $\tau = 3, 4, 5$ fm/c.
As the slab is at rest ($v^z =0$) in the Cartesian coordinate, which corresponds to 
$v^\eta < 0$ in Milne coordinates.
For $\eta > 0$, a rarefaction wave starts at the edge of the slab (i.e at the discontinuity) 
and propagates within the slab with a velocity $c_s$. Also a shock starts at the discontinuity and moves 
into the vacuum with the speed of light. Such features are also observed at $\eta < 0$ (not shown here).
For instance at $\tau = 3$ fm/c, we find $v^z =0$ for $|\eta| \leq 0.5$ and thus $v^\eta < 0$.
The rarefaction wave has then spread outside the slab from $0.5 < |\eta| < 2.6$, where $|\eta|=2.6$
correspond to the boundary of the vacuum. The fluid expands outward with increasing velocity 
$v^\eta$ till it reaches the boundary of the vacuum. The wave velocity $v^z$ rapidly increases
outward and approaches the speed of light $v_z/c \approx 1$ at the vacuum. We find that our numerical
results are in perfect agreement with the Riemann simple wave solution for all rapidity $\eta$
and all proper times $\tau$.

\begin{figure}[t]
\includegraphics[width=\linewidth]{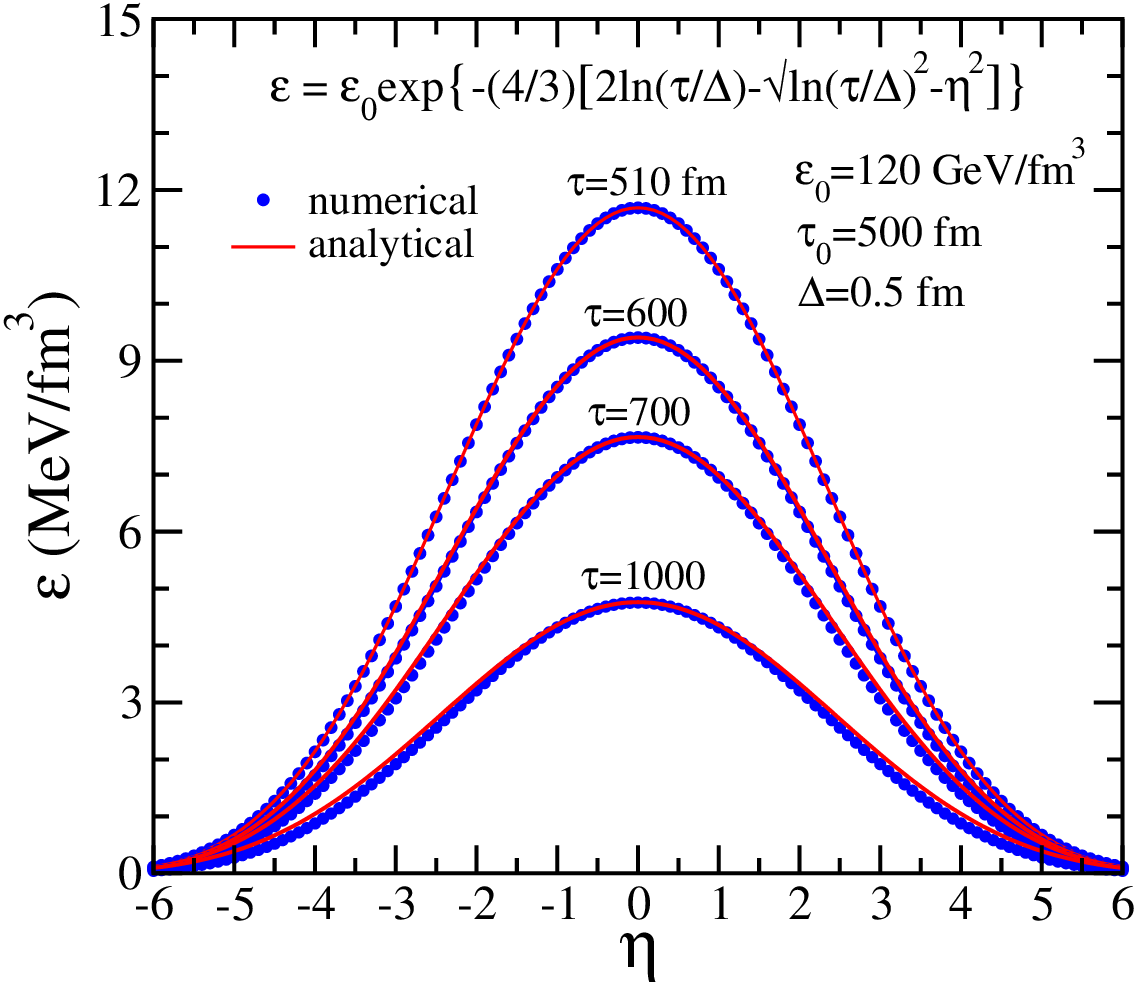}
\caption{Comparison between the analytical solutions (lines) and the numerical results
(circles) for the rapidity dependence of energy density $\epsilon$ at proper times 
$\tau = 510, 600, 700, 1000$ fm/c for the initial condition of Eq. (\ref{LK:eq})
at time $\tau_0 = 500$ fm/c.}
\label{fig:LK}
\end{figure}

With increasing time, the two rarefaction waves, traversing inwards from positive and negative
rapidity sides of the slab, will eventually reach the center $\eta=0$ of the slab and overlap.
In practice, from this time onwards, Riemann solution cannot be applied, and
the Landau-Khalatnikov solutions starts to be applicable. The Landau-Khalatnikov 
solution at the later times describe evolution of matter in the overlap region of the slab.
At the asymptotic times $\tau \gg \Delta$, the Landau-Khalatnikov solution can be expressed as 
\begin{align}\label{LK:eq}
\epsilon = \epsilon_0 \left[ - \frac{4}{3} \left\{ 2 \ln \left(\frac{\tau}{\Delta}\right)
- \sqrt{ \ln\left(\frac{\tau}{\Delta}\right)^2 - \eta^2} \right\} \right],
\end{align}
with $v^\eta = 0$. Here $\Delta$ is the thickness of the slab where the two ingoing rarefaction 
waves overlap.

In the numerical simulation, the initial energy distribution are obtained from Eq. (\ref{LK:eq}) 
corresponding to initial values of time $\tau_0 = 500$ fm/c, energy density 
$\epsilon_0 =120$ GeV/fm$^3$ and slab thickness $\Delta = 0.5$ fm. Figure \ref{fig:LK}
shows the comparison between numerical results and the Landau-Khalatnikov asymptotic
solution. The calculations are in good agreement with the analytical results up to large times
especially around central rapidity region. At large rapidity the deviations from the asymptotic
value may suggest that the rarefaction (overlapping) wave is mostly confined around the center 
region thus making the Landau-Khalatnikov asymptotic results invalid at large $|\eta|$.

\subsection{Initial conditions for non-boost-invariant expansion}

The initial conditions for our (1+1)D non-boost-invariant expansion of the viscous fluid at the initial
proper time $\tau_0$ is defined by the three quantities, viz. 
$\epsilon(\tau_0,\eta)$, $v_\eta(\tau_0,\eta)$, $\pi^{xx}(\tau_0,\eta)$. 
In our simulation we have adopted $\tau_0 = 0.4$ fm/c at which the initial energy density is 
taken as \cite{Schenke:2011bn}
\begin{align}\label{engd:eq}
\epsilon(\tau_0,\eta) = \epsilon_0 \exp\left[ - \frac{\left(|\eta| - \Delta\eta\right)^2}{2\sigma_\eta^2}
\theta\left(|\eta| - \Delta\eta\right) \right] .
\end{align}
This profile consists of a flat distribution about midrapidity of width $2\Delta\eta$  and two-smoothly 
connected Gaussian tails of half-width $\sigma_\eta$. The parameters 
$\epsilon_0$ and ($\Delta\eta,\sigma_\eta$)  are adjusted to reproduce the absolute magnitude and width
of the final rapidity distribution of mesons measured by BRAHMS \cite{Bearden:2004yx} 
in central Au+Au collisions at the RHIC energy of $\sqrt{s_{NN}}=200$ GeV.  
The initial values of the longitudinal velocity profile is taken as 
boost-invariant, and the viscous stress tensor as isotropic:
\begin{align}\label{velpi:eq}
v_\eta(\tau_0,\eta) = 0, ~~~~ \pi^{mn}(\tau_0,\eta) = 0.
\end{align}

The hydrodynamic evolution is continued until each fluid cell reaches a decoupling 
temperature of $T_{\rm dec} = 150$ MeV. 
We consider only direct pion and kaon production and do not include their formation from
resonance decays. To account for the latter contribution, we follow the prescription 
of \cite{Bozek:2007qt} by noting that, since $\sim 75\%$ of pions originate from 
resonance decays \cite{Torrieri:2004zz}, the rapidity distribution of Eq. (\ref{rapidity:eq}) 
is multiplied by a factor of four.

The equation of state (EoS) influences the longitudinal expansion of the fluid and
the two-particle correlations.
In this work, the effects of EoS on the correlators have studied by 
employing both a conformal QGP fluid with the thermodynamic pressure 
$p = \epsilon/3$, and the s95p-PCE EoS \cite{Huovinen:2009yb} which is obtained
from fits to lattice data for crossover transition and matches to
a realistic hadron resonance gas model at low temperatures $T$,
with partial chemical equilibrium (PCE) of the hadrons at 
temperatures below $T_{\rm PCE} \approx 165$ MeV.
Unless otherwise mentioned, the shear relaxation time in Eqs. (\ref{IS:eq}) and 
(\ref{SOSHEAR}) is set at $\tau_\pi = 5\eta_v/4p$ corresponding to 
$\tau_\pi = 5\eta_v/(sT)$ in the conformal fluid.

\label{parameter}
\begin{table}
\caption{The parameters of the initial energy distribution $\epsilon_0, \sigma_\eta$
with $\Delta\eta =0.6$ corresponding to Eq. (\ref{engd:eq}), 
that reproduce the final pion rapidity distribution in $0-5\%$ central Au+Au 
collisions at $\sqrt{s_{NN}}=200$ GeV. The results are in ideal and viscous hydrodynamic 
evolution in the MIS theory with EoS for conformal fluid, 
and that for lattice EoS are shown in braces. The last column gives the lifetime 
of the fluid at a freeze-out temperature of $T_{\rm dec} = 150$ MeV.} 

\begin{tabular}{l l l l l l l l}\hline\hline
$\eta_v/s$   &\hfil & $\epsilon_0$ (GeV/fm$^3$) &\hfil & $\sigma_\eta$  &\hfil & $\tau_f$ (fm/c) \\ \hline

0      &\hfil  & 200 (39.0)  &\hfil & 0.9   &\hfil  & 16.40 (15.84)  \\
0.08   &\hfil  & 142 (27.1)  &\hfil & 1.0   &\hfil  & 16.08 (15.36)  \\
0.24   &\hfil  & 82  (17.5)  &\hfil & 1.4   &\hfil  & 14.88 (14.16)  \\ \hline\hline

\end{tabular}
\end{table}

\begin{figure}[b]
\includegraphics[width=\linewidth]{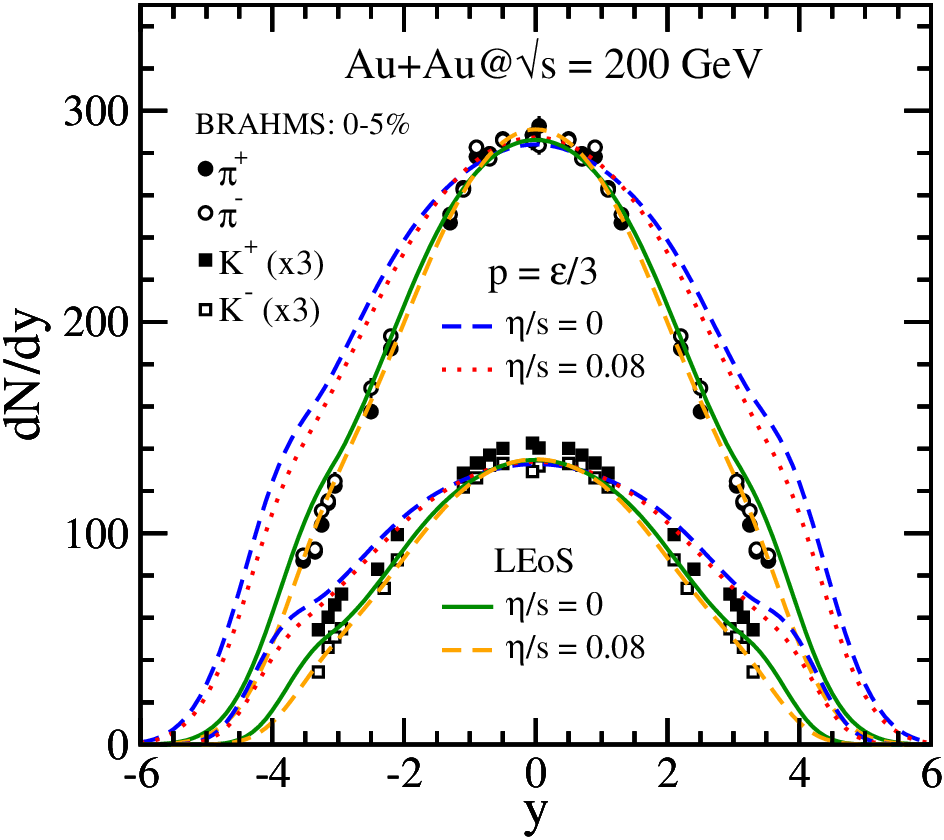}
\caption{Rapidity distribution of $\pi^\pm$ and $K^\pm$ in $0-5\%$ central Au+Au collisions 
at $\sqrt{s_{NN}} = 200$ GeV. The symbols represent the BRAHMS data \cite{Bearden:2004yx},
and the lines correspond to non-boost-invariant hydrodynamic calculations for the conformal fluid and 
lattice equation of state at the shear viscosity to entropy density ratio of $\eta_v/s =0$ and 0.08
in the MIS approach.}
\label{fig:dNdy}
\end{figure}

Figure \ref{fig:dNdy} shows the rapidity distribution of pions and kaons 
in our (1+1)D non-boost-invariant hydrodynamic model as compared with the $5\%$ most central
Au+Au collision data from BRAHMS \cite{Bearden:2004yx}.
The parameters ($\epsilon_0, \Delta\eta,\sigma_\eta$) obtained by fitting the 
data for pions at midrapidity are listed in Table I. The stiff conformal
EoS induces an accelerated longitudinal flow with a much wider dN/dy as compared to the data.
In fact, this EoS fails to reproduce the data at large rapidities for any combination
of the parameters (or with varying $T_{\rm dec}$). In contrast, the softening produced
due to deconfinement transition in the lattice EoS gives a smaller longitudinal
pressure gradients and leads to a better agreement with rapidity distributions.
As compared to ideal-hydrodynamics, the second-order viscous hydrodynamics slows down
the expansion of the fluid, and thus requires a smaller and wider initial energy density 
distribution (see Table I) to be compatible with the final meson rapidity distribution.

\begin{figure}[t]
\includegraphics[width=\linewidth]{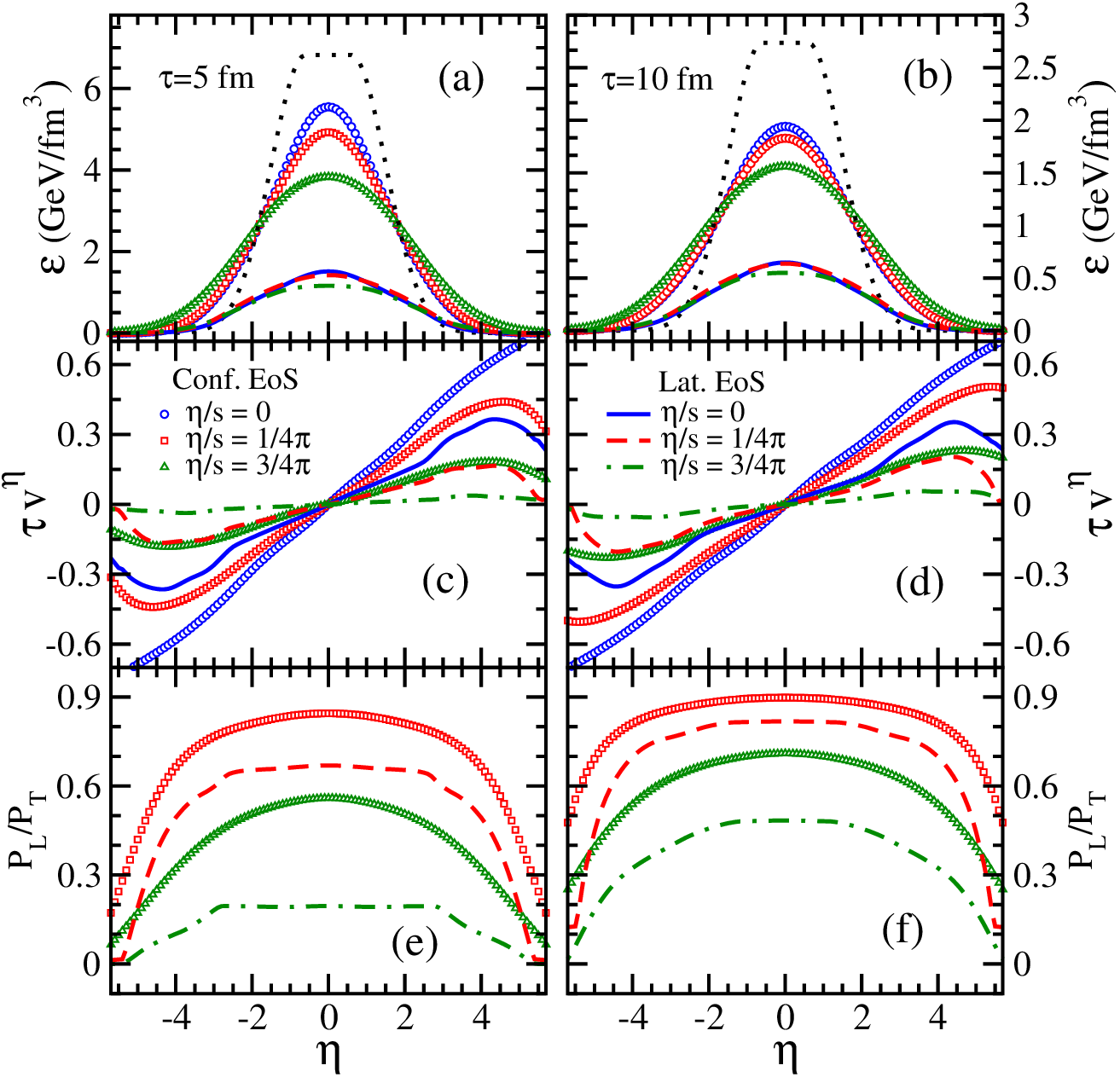}
\caption{Space-time rapidity dependence of (a)-(b) energy density, (c)-(d) longitudinal velocity
(scaled by time), and (e)-(f) longitudinal to transverse pressure ratio $P_L/P_T$, 
in the hydrodynamic calculations at proper times of $\tau = 5$ fm (left panels) and 
10 fm (right panels). The results are for ultra-relativistic gas (symbols) and 
lattice (lines) equations of state with $\eta_v/s = 0, 1/4\pi, 3/4\pi$ in the MIS theory. 
The black dotted line is the Bjorken scaling solution $\epsilon/\epsilon_0 \sim (\tau_0/\tau)^{4/3}$.}
\label{fig:evp}
\end{figure}

In Fig. \ref{fig:evp} we present the space-time rapidity dependence of energy density $\epsilon$,
longitudinal flow velocity $v_\eta$, and the ratio of longitudinal and transverse pressure,
$P_L/P_T = (p_0-\pi_0/2)/(p_0+\pi_0/2)$ at the proper 
times of $\tau = 5$ and 10 fm obtained in our non-boost-invariant model.
The energy density in the perfect-fluid conformal hydrodynamics (blue circles in Fig. \ref{fig:evp}(a)-(b)) 
shows a decreasing flat region at midrapidity with increasing time as compared to initial profile.
In contrast to boost-invariant expansion, the stronger longitudinal expansion due to larger pressure
gradients in the non-boost-invariant case transfer energy faster to larger rapidities. Indeed, 
in the Bjorken scaling solution for a perfect-fluid, the energy density 
$\epsilon(\tau) = \epsilon(\tau_0) (\tau_0/\tau)^{4/3}$ (black dotted line) is seen to lie above (below)
than that in the (1+1)D case at midrapidity (large rapidities). Thus, in general, a Bjorken expansion
would underestimate the cooling of the system. The discrepancies become larger with time as can be seen 
in Fig. \ref{fig:evp}(b) at $\tau =10$ fm/c. The inclusion of viscosity slows down the expansion 
and thereby the cooling of the system. As a result, the energy density
distribution for $\eta_v/s =0.08$ (red squares) and $\eta_v/s =0.24$ (green triangles) for the
ultra-relativistic gas becomes increasingly comparable to the perfect fluid case, inspite
of smaller initial energy values in the dissipative hydrodynamics (see Table I).
For the softer lattice EoS, the differences in the energy densities for various
$\eta_v/s$ (lines in Fig. \ref{fig:evp}(a)-(b)) become increasingly smaller with increasing 
$\tau > \tau_0$.  

Figure \ref{fig:evp}(c)-(d) shows that the longitudinal flow velocity $v_\eta$ distribution
(multiplied by the corresponding proper time) rapidly increases with rapidity in the ideal-fluid dynamics.
Although we initialized the fluid with a boost-invariance value at all rapidities, i.e. 
$v_\eta(\eta,\tau_0) =0$, the longitudinal pressure gradients quickly accelerate the fluid 
and breaks the longitudinal boost-invariance at $\tau > \tau_0$. In our second-order viscous
evolution, viscosity restricts the pressure gradients and reduces the increase of $v_\eta$ with
$\eta$. At large rapidities, the smaller pressure (and temperature) enhances the time
$\tau_\pi = 5\eta_v/4p$ for the system to relax towards equilibrium. As a consequence,
the larger viscous corrections here decelerates the expansion and eventually overcomes 
the acceleration from pressure gradients. This causes $v_\eta$ to approach the Bjorken 
limit, and beyond this rapidity the second-order viscous hydrodynamics becomes questionable.
With increased $\eta_v/s =0.2$, the stronger viscous effects drive the system toward this
unphysical behavior at an earlier rapidity. As the longitudinal flow velocity build-up
with increasing evolution time, at the later time of $\tau=10$ fm/c, its appearance is 
delayed to larger rapidity value (see Fig. \ref{fig:evp}). 
Compared to the stiff EoS, $p=\epsilon/3$, the lattice EoS injects a smaller pressure gradient 
resulting in a smaller deviation from the Bjorken flow profile, especially at large $\eta_v/s$. 

The pressure anisotropy $P_L/P_T$ (\ref{fig:evp}(e)-(d)) shows marked deviation from the 
isotropic initial pressure configuration of $P_L/P_T=1$. As the shear stress 
tensor $\pi^{\eta\eta}$ gradually build-up with time and later decreases slightly, the anisotropy
is larger at $\tau= 5$ fm/c than at 10 fm/c. At large rapidities, $\pi^{\eta\eta}$ becomes comparable 
to the thermodynamic pressure, hence $P_L/P_T$ rapidly decreases and can eventually become 
negative. Although an increase in $\eta_v/s$ leads to a smaller $P_L/P_T$ at midrapidity, a 
somewhat wider initial energy distribution (see Table I) prevents an early appearance of this unphysical 
region at any given time. As expected, the dissipative effects are more pronounced in the
lattice EoS and results in larger pressure anisotropy.

The large space-time variation of the flow and pressure anisotropy, as found here in a finite 
fluid system, should have important effects on the two-particle rapidity correlations arising
from the propagation of thermal noise.

\subsection{Single thermal fluctuation on top of (1+1)D viscous expanding medium}

For a clear understanding of the evolution of hydrodynamic fluctuations and the resulting rapidity 
correlations induced by thermal fluctuations at all space-times, 
it is instructive to focus first
on the evolution of one static thermal perturbation. In particular, we consider a static
Gaussian thermal fluctuation induced at ($\eta_0,\tau_0$) on top of non-boost-invariant expanding medium:
\begin{align}\label{deltaT}
\delta T (\eta,\tau_0) =& ~ T(\eta_0,\tau_0) \frac{\kappa}{(2\pi\sigma)^{1/2}} 
{\rm exp}\left[-(\eta - \eta_0)^2/2\sigma^2 \right], \nonumber\\
\delta u^\eta (\eta,\tau_0) =& ~\delta \pi^{\eta\eta}(\eta,\tau_0)  = 0 .
\end{align}
At the initial time $\tau_0 = 0.4$ fm/c, the perturbation is induced at
the fluid rapidity $\eta_0$ and has a Gaussian width parameter 
$\sigma =0.2$ and amplitude $\kappa = 0.1 \sqrt{2\pi\sigma}$. 

The perturbation travels in opposite directions such that in the local rest frame of the 
background fluid, the speed of propagation is the sound velocity $c_s = \pm (dz/dt)_\mathrm{LRF}$. 
Writing this covariantly one obtains 
$u_0^\mu \varepsilon_{\mu\alpha} dx^\alpha = \pm c_s u_{0 \mu} dx^\mu $,
where $\varepsilon^{\mu\nu}$ is the totally
antisymmetric tensor of rank 2. Noting that in $(\tau,\eta)$ coordinates $\varepsilon_{\tau\eta} = \tau$, 
the equation of motion of the perturbation peak is found to be,
\begin{equation}\label{cs_evol}
\tau \frac{d\eta}{d\tau} = \frac{\pm c_s + \tau v_{\eta_0}}{1 \pm c_s \tau v_{\eta_0}}
\end{equation}
In case of Bjorken expansion where $v_{\eta_0} = 0$, the above expression has the simple solution
$\eta = \eta_0 \pm c_s \log(\tau/\tau_0)$. For the (1+1)D expansion, Eq. (\ref{cs_evol}) has to be integrated numerically
as $v_{\eta_0}(\tau,\eta)$ is not known analytically. It is important to note that unlike the Bjorken case,
where the extent of propagation of perturbations is independent of the dissipative equations considered,
in the (1+1)D case the shear stress tensor controls the width of sound cone by determining
the background flow profile $v_{\eta_0}$ seen in Fig. \ref{fig:evp}.

In addition to influencing the trajectories of perturbations, an expanding fluid 
also leads to diffusion of the propagating disturbance. For example, in an ideal fluid at rest,
a perturbation would propagate unattenuated in opposite directions at the speed of sound, whereas,
in a Bjorken expansion (even in the ideal limit) the disturbance broadens and dampens during propagation.
This is attributed to the non-linear dispersion relation for wave propagation on top of Bjorken
expansion, $\omega(k) = i(1-c_s^2)/2 \pm \sqrt{c_s^2 k^2 - (1-c_s^2)^2/4}$. In the (1+1)D expansion, 
the dispersion relation becomes complicated
via dependences on space-time and has to be obtained numerically.
The propagation of temperature disturbance will induce perturbation in velocity $\delta u^\eta$ 
and shear pressure tensor $\delta\pi$ at later times $\tau > \tau_0$.
 
The rapidity distribution of the correlations induced by these fluctuations can be explored 
via the equal-time rapidity correlation,
\begin{align}\label{scor}
C_{\Delta X,\Delta Y}(\tau,\Delta\eta;\eta_0) = \int d\eta' \Delta X(\eta',\tau) \Delta Y(\eta'+\Delta\eta,\tau) ,
\end{align}
where $\eta_0$ is the initial position of the disturbance at time $\tau_0$ and $(\Delta X,\Delta Y)$ refer to the
the normalized fluctuations $\Delta T = \delta T(\eta,\tau)/T_0(\eta,\tau)$, 
$\Delta u^\eta = \tau \delta u^\eta$ and $\Delta\pi = \delta\pi/(\epsilon_0+p_0)$.
Due to explicit
dependence of the background evolution on space-time rapidity $\eta$, the above correlator 
would depend on the initial $\eta_0$ where the perturbation is introduced.
This is to be contrasted with the Bjorken expansion where the translational invariance 
(in $\eta$ direction) of the background flow implies that $C_{\Delta X, \Delta Y}$ does not 
depend on the initial rapidity position of the perturbation \cite{Shi:2014kta}.

\begin{figure}[t]
\includegraphics[width=\linewidth]{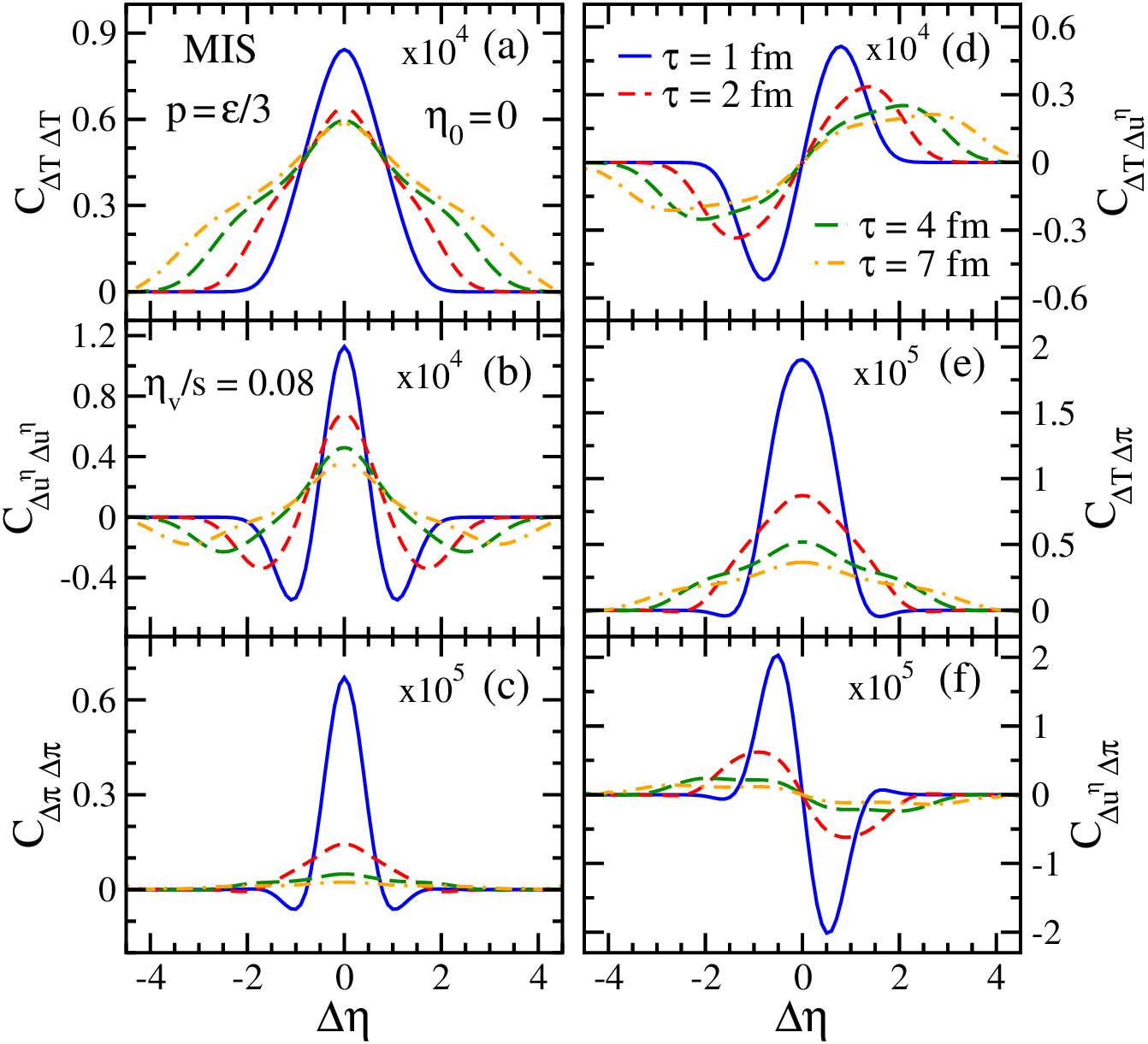}
\caption{Equal-time longitudinal rapidity correlations $C_{\Delta X,\Delta Y}(\tau,\Delta\eta;\eta_0)$ with
$(\Delta X,\Delta Y) \equiv [\delta T/T_0, \tau\delta u^\eta, \delta\pi/(\epsilon_0+p_0)]$ computed as 
a function of space-time rapidity separation $\Delta\eta$ at various later times 
from a thermal perturbation at an initial time of $\tau_0 = 0.4$ fm/c and at rapidity $\eta_0=0$.
The perturbation is induced on top of non-boost-invariant background hydrodynamic 
flow in the MIS theory using an ultra-relativistic gas EoS with $\eta_v/s = 1/4\pi$. 
The correlators in each panel are vertically scaled by a representative value.}
\label{fig:statcor}
\end{figure}

In Fig. \ref{fig:statcor}, we present rapidity correlations 
at later times arising due to a static initial thermal perturbation 
induced at the center $\eta_0=0$ of the background fluid undergoing (1+1)D hydrodynamic expansion 
in the MIS theory. We consider a $p=\epsilon/3$ EoS and $\eta_v/s = 0.08$; the other initial and
freeze-out conditions of the background are given in Table I. Figure \ref{fig:statcor}(a) shows that 
at early times the temperature-temperature rapidity correlation $C_{\Delta T\Delta T}$
has a large and narrow peak at $\Delta\eta=0$ due to self-correlations. With the expansion 
of the background fluid, the amplitude of the peak decreases with time 
as it spreads over a large rapidity separation leading to long-range
rapidity correlations till the freeze-out of the system is reached at $\tau_f \approx 16$ fm/c.
The extent of the rapidity correlation at any given time is bounded by the maximum
distance travelled by the sound wave, namely the sound horizon, which can be obtained by solving  
Eq. (\ref{cs_evol}). In contrast, rapidity correlation 
from ripples on top of a boost-invariant ideal background fluid was shown \cite{Shi:2014kta}
to generate a sharp peak at $\Delta\eta \approx 0$, followed by a relatively flat region at
intermediate $\Delta\eta$, and a much smaller peak from the sound horizon at large $\Delta\eta$. 
Thus, in the present (1+1)D viscous hydrodynamic expansion, the much broader rapidity correlation 
(with negligibly small second peak) that persists even at late times can be attributed to the 
interplay of nonzero background fluid velocity $v_{\eta_0}(\tau,\eta)$ and viscous damping in the MIS theory.

Figure \ref{fig:statcor}(b) shows the time-evolution of velocity-velocity rapidity correlation 
$C_{\Delta u^\eta \Delta u^\eta}$. Starting with an initial value of $\delta u^\eta =0$, the velocity
perturbations and correlations at first build-up with time at about zero rapidity 
separation and then decreases later when the perturbation spreads to large rapidities.
The negative correlations seen at larger rapidity separations are essentially due to $\delta u^\eta$
having opposite signs along positive and negative directions relative to the initial position
of perturbation. Careful examination of Figs. \ref{fig:statcor}(a), (b) 
shows that the minima in this negative correlations are produced at the sound horizon corresponding 
to the ``second peak" in the temperature-temperature correlation.

The pressure-pressure rapidity correlation due to shear $C_{\Delta\pi \Delta\pi}$ shown 
in Fig. \ref{fig:statcor}(c), exhibits a similar peaked structure as seen for temperature-temperature 
correlations. However, the magnitude of this correlation is much smaller and does not spread 
much in rapidity separation with increasing time. Note that in the present initialization
of temperature perturbation (instead of velocity or shear-pressure perturbations), the 
magnitude of $C_{\Delta T\Delta T}$ dominates and it is about two orders of magnitude 
larger than $C_{\Delta u^\eta \Delta u^\eta}$. 

On the other hand, the rapidity correlations $C_{\Delta T \Delta u^\eta}$ 
and $C_{\Delta u^\eta \Delta\pi}$ [see Figs. \ref{fig:statcor}(d), (f)] 
are odd functions of $\Delta\eta$ and thus the correlations vanish at $\Delta\eta =0$. 
Moreover, the ``cross" correlations follow $C_{\Delta X, \Delta Y} = -C_{\Delta X, \Delta Y}$. 
The structure of the $C_{\Delta T \Delta \pi}$ correlator in Fig. \ref{fig:statcor}(e) can
be easily understood from the $C_{\Delta T \Delta T}$ and $C_{\Delta \pi \Delta \pi}$ correlations.

\begin{figure}[t]
\includegraphics[width=\linewidth]{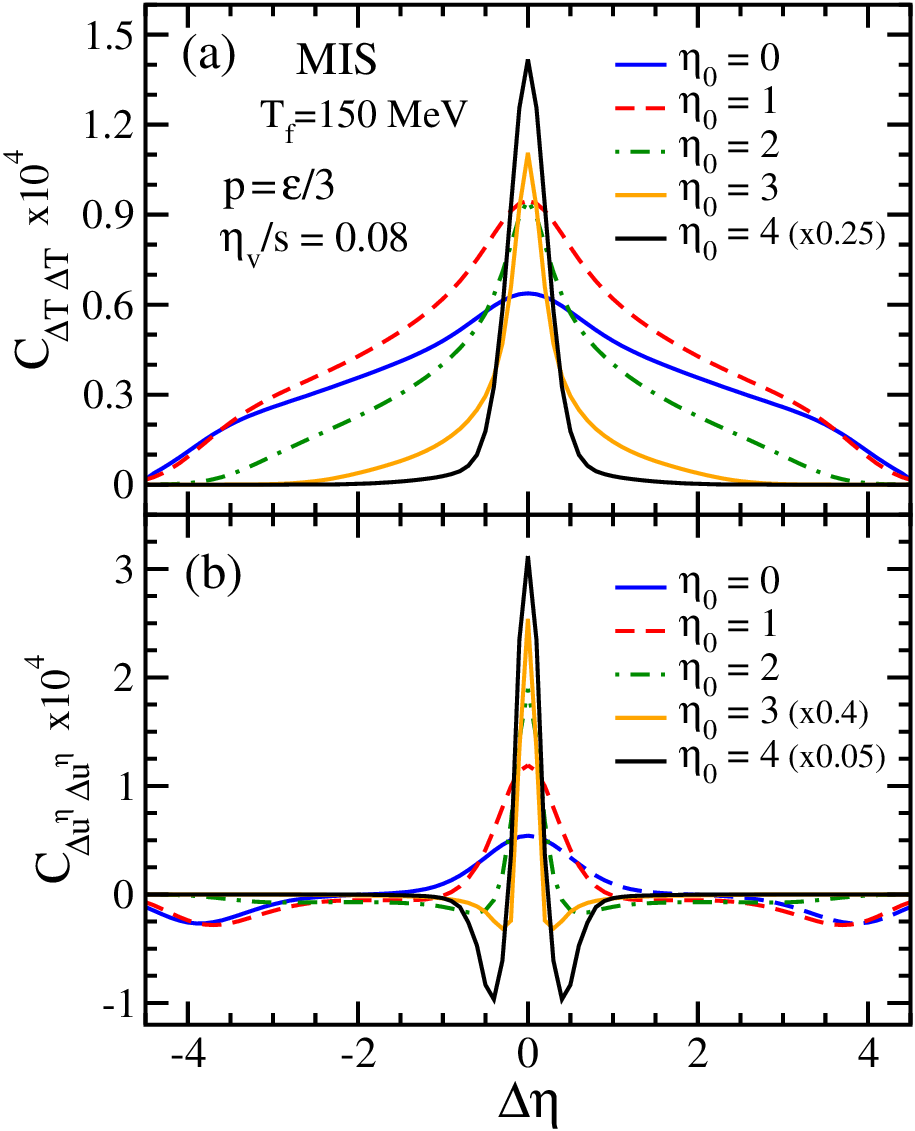}
\caption{Temperature-temperature and velocity-velocity rapidity correlations
as a function of rapidity separation $\Delta\eta$ arising from perturbations at various
initial rapidities $\eta_0$ and computed at a freeze-out hypersurface $T_{\rm dec} = 150$ MeV.
The perturbations are induced on top of non-boost-invariant background hydrodynamic
flow in the MIS theory for an ultra-relativistic gas EoS with $\eta_v/s = 1/4\pi$.
The correlators for large $\eta_0$ are scaled vertically by the values shown within braces.}
\label{fig:statcor_frz}
\end{figure}

In Figs. \ref{fig:statcor_frz}(a), (b) we show the correlations between temperature-temperature 
and velocity-velocity at the freeze-out hypersurface $T(\tau_f,\eta) = T_\mathrm{dec}$, induced 
by a single temperature perturbation placed at various 
initial rapidity values $\eta_0$. Accordingly, we now use the definition of the correlator 
$C_{\Delta X,\Delta Y}(T_\mathrm{dec},\Delta\eta;\eta_0) = \int d\eta' \Delta X(\eta',T_\mathrm{dec}) \Delta Y(\eta'+\Delta\eta,T_\mathrm{dec})$. 
For perturbation introduced at a large rapidity, we find the self-correlations to increase
and the long-range correlations to decrease. This is because the fluid cells at large 
rapidities (having small initial temperatures) freeze-out at early times \cite{Florkowski:2016kjj} 
and thus one of the perturbation peaks which propagates along the background flow 
reaches the freeze-out hypersurface quickly and is effectively undamped.
Consequently, the self-correlations which are essentially squares of the peak values, increase with $\eta_0$
of the initial perturbation. However, the long-range correlation which depends on the product of the two peaks
decrease as the other peak which travels opposite to the background fluid takes substantially longer to reach the 
freeze-out surface and is almost fully damped; see Fig. \ref{fig:statcor_frz}(a). 
As seen in Fig. \ref{fig:statcor_frz}(b), the rise in self-correlations with $\eta_0$ is found 
to be more for the velocity-velocity correlator due to the pronounced background acceleration of 
the fluid at large $\eta$, which leads to build up of the velocity of the travelling perturbation.

\subsection{Thermal noise correlations on top of (1+1)D viscous expanding medium}

In this section, we shall explore longitudinal rapidity correlations induced by thermal
fluctuations in the non-boost-invariant (1+1)D expansion of the background medium.
These fluctuations, which act as source terms for linearized hydrodynamic equations
are correlated over short length-scales, and accordingly they generate singularities in 
the correlators for hydrodynamic variables at zero rapidity separation and at the sound 
horizons. While for Bjorken expansion within the Navier-Stokes theory, 
the correlators can be analytically decomposed into regular and singular parts \cite{Chattopadhyay:2017rgh}, 
in the second-order MIS and CE theories such analytic separation is not plausible. 
In Eq. (\ref{FRapCor:eq}) for the two-particle rapidity 
correlations calculated at freeze-out, the singularities get smeared out by the
coefficients ${\cal F}_X$ [with $X \equiv (T_1, u_1^\eta, \pi_1)$], thereby 
allowing for a smooth presentation. In order to explore at various times
$\tau \leq \tau_f$, the equal-time longitudinal rapidity correlation arising from thermal noise, 
we consider a Gaussian convoluted correlation    
\begin{align}\label{ncor}
\langle C_{X,Y}(\Delta\eta,\eta;\tau) \rangle = 
& \int d(\Delta\eta') \langle X(\eta,\tau) Y(\eta+\Delta\eta',\tau) \rangle \nonumber \\
& \times {\rm exp} \left[-(\Delta\eta - \Delta\eta')^2/2\sigma_{\small \Delta\eta}^2 \right],
\end{align}
where $(X,Y) \equiv (\delta T, \delta u^\eta, \delta\pi )$ refer to the usual perturbations in
the event. An averaging $\langle \cdots \rangle$ has been performed over many fluctuating 
events that evolve on top of background hydrodynamics; the initial conditions for the latter is 
given in Table I.
Note that although the qualitative nature of the correlations
are insensitive to the smearing function, whose width we have taken as $\sigma_{\Delta\eta} = 0.4$, the 
the magnitude and spread of the peaks depend on the latter.

\begin{figure}[t]
\includegraphics[width=\linewidth]{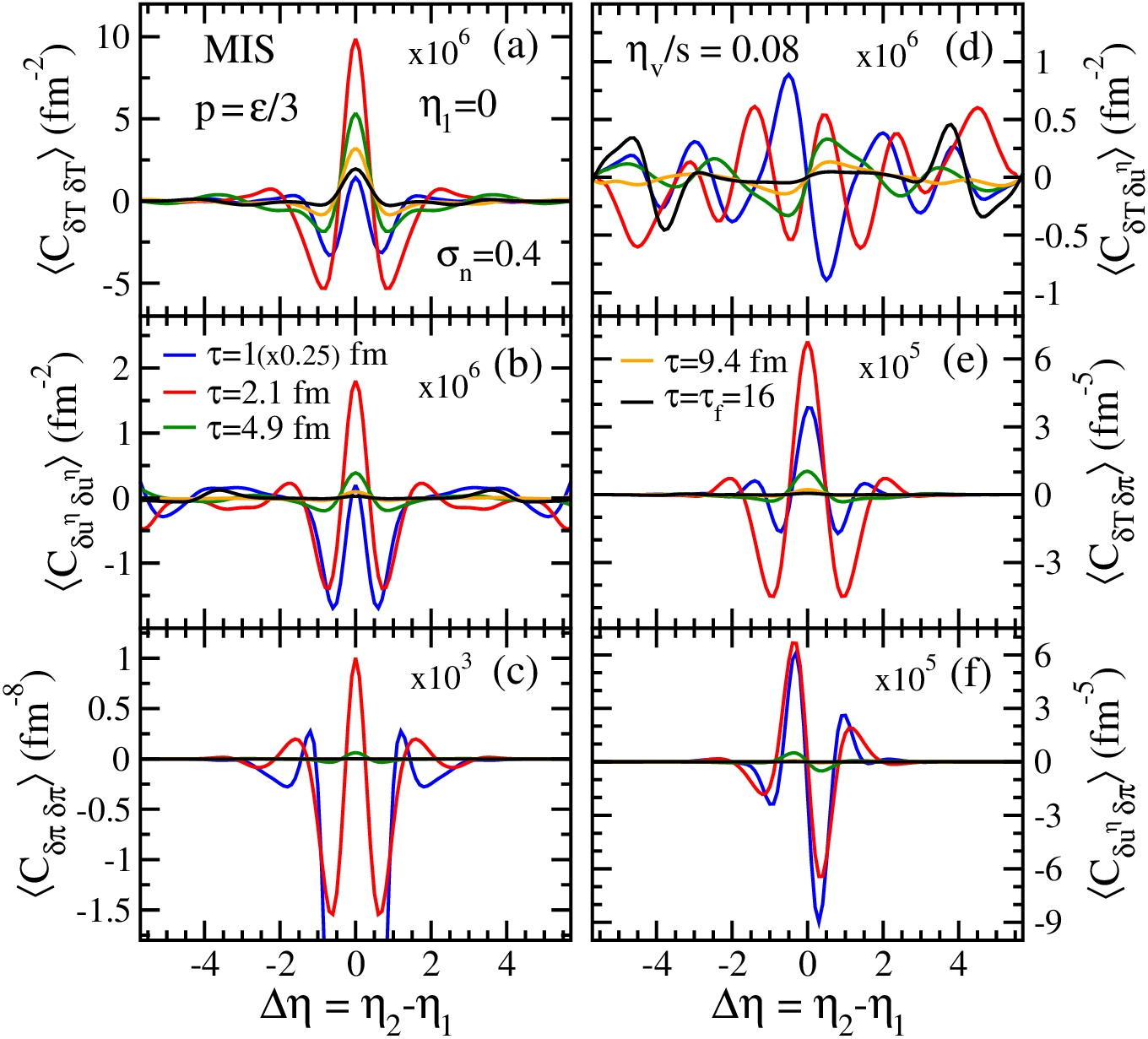}
\caption{Event averaged equal-time longitudinal rapidity correlations 
$\langle C_{X,Y}(\Delta\eta,\eta) \rangle$ with 
$(X,Y) \equiv (\delta T, \delta u^\eta, \delta\pi)$ computed as a function of
space-time rapidity separation $\Delta\eta= \eta_1-\eta_2$ about $\eta_1=0$ 
at various times due to thermal noise perturbations 
on top of non-boost-invariant background flow. The results are in the 
M\"uller-Israel-Stewart (MIS) theory for ultra-relativistic gas EoS with 
$\eta_v/s = 1/4\pi$. The correlators in each panel are scaled vertically by values given; 
the correlations at $\tau = 1$ fm/c are further scaled by 0.25 for clarity.} 
\label{fig:noiscor}
\end{figure}

\begin{figure}[t]
\includegraphics[width=\linewidth]{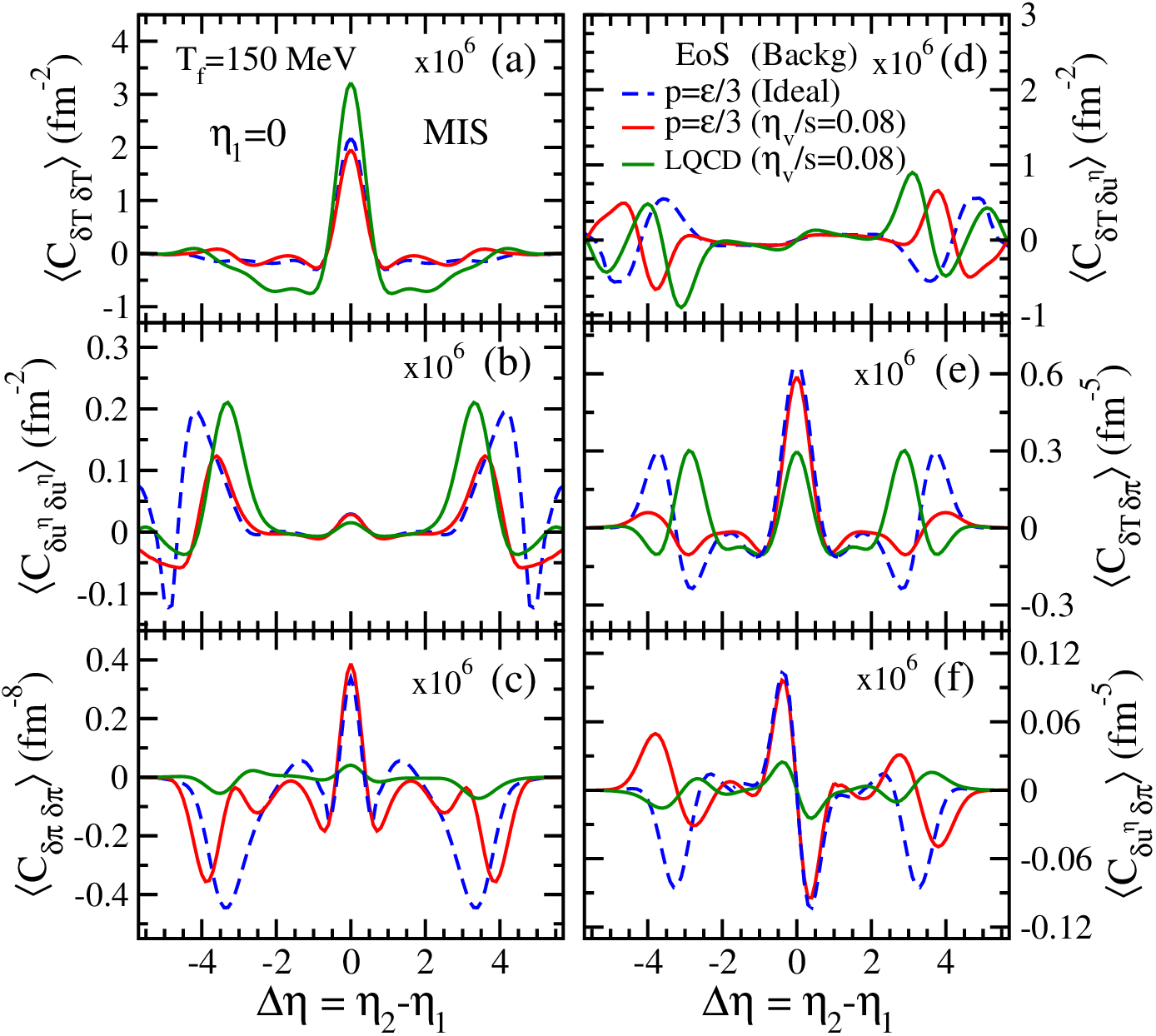}
\caption{Event averaged rapidity correlations $\langle C_{X,Y}(\Delta\eta,\eta) \rangle$ 
with $(X,Y) \equiv (\delta T, \delta u^\eta, \delta\pi)$ at midrapidity
from thermal noise at the freeze-out temperature $T_{\rm dec} = 150$ MeV.
The results are in the MIS formalism with $\eta_v/s = 1/4\pi$ in the average
and thermal noise evolution for $p=\epsilon/3$ EoS (red solid line), lattice EoS (green solid line) and 
an ideal background evolution with $p=\epsilon/3$ EoS (blue dashed line).
The initial and final conditions for the background evolution are given in Table I. 
All the correlations are scaled vertically by $10^6$.}
\label{fig:frz_MICE}
\end{figure}

Figure \ref{fig:noiscor} displays equal-time rapidity correlations for various fluctuations 
arising from thermal noise on top of (1+1)D hydrodynamic expansion in the MIS theory
with $\eta_v/s = 0.08$ in both the background and noise evolution equations. 
Thermal fluctuations at each spatial point and during the entire evolution of the fluid
produce short-range temperature-temperature correlation peaked at zero rapidity separation,
see Fig. \ref{fig:noiscor}(a). In contrast to correlation from an initial perturbation 
[see Fig. \ref{fig:statcor}(a)], a much narrower peak is seen in thermal noise. 
The appreciable negative correlations at small rapidity separation is due to the 
second-derivative of the delta function arising from the 
noise term in momentum conservation equation. In fact, the magnitude of the peaks
and troughs are dominated by the singularities that occur at $\Delta \eta =0$ due to 
self-correlations and at sound horizons.  
This lead to non-monotonous structures in the correlations induced by thermal noise 
at large $\Delta\eta$ 
in contrast to that seen from a single perturbation. At later times, the expansion of 
the fluid causes the peak values to decrease and the correlations to spread somewhat 
farther in rapidity separations.

The velocity-velocity and shear pressure-pressure rapidity correlations shown
in Figs. \ref{fig:noiscor}(b), (c) also give pronounced negative correlations from the singularities
at small $\Delta\eta$. While the $\langle C_{\delta u^\eta,\delta u^\eta}\rangle$ correlation give nontrivial 
structures about the sound horizon, the $\langle C_{\delta\pi,\delta\pi}\rangle$ correlation essentially
has a small magnitude and rapidly damp at larger $\Delta\eta$. Consequently, the cross
correlations $\langle C_{\delta T,\delta\pi} \rangle$  
and $\langle C_{\delta u^\eta,\delta\pi} \rangle$  have negligible values at large rapidities 
and contribute minimally to the final two-particle rapidity correlations.
By inspection of Figs. \ref{fig:statcor} and \ref{fig:noiscor} it is clearly evident that compared
to an induced perturbation, the realistic hydrodynamic fluctuations in (1+1)D expansion 
generate rich structures at short and long range two-particle rapidity correlations.

To gauge the importance of underlying flow and viscous damping, we compare 
in Fig. \ref{fig:frz_MICE} the rapidity correlations 
$\langle C_{X,Y}(\Delta\eta,\eta) \rangle$ at the freeze-out hypersurface 
corresponding to $T_{\rm dec} = 150$ MeV in the MIS theory at $\eta_v/s = 1/4\pi$ (red solid lines)
and also for ideal background hydrodynamic evolution (blue dashed line) for
ultra-relativistic gas EoS.
In absence of viscous damping larger peaks and troughs can be seen at small
$\Delta\eta$. Moreover, the fluctuations travel over large rapidity separation  
and generate distinct structures about the sound horizon. 
We also present correlations computed for a lattice EoS in the MIS theory 
at $\eta_v/s = 1/4\pi$ (green solid lines).
The smaller sound velocity near the deconfinement transition slows down the expansion of
the background fluid (see Fig. \ref{fig:evp}) as well as limits the spatial extent of the 
sound horizon. These lead to sharp peaks from self-correlation and large and broad negative
correlations from the singularities at $\Delta\eta \approx 0$ and sound horizon. 
In fact, the total correlation in the lattice is dominated by the temperature-temperature
correlations.

\subsection{Two-particle rapidity correlations in (1+1)D expanding medium}

\begin{figure}[t]
\includegraphics[width=\linewidth]{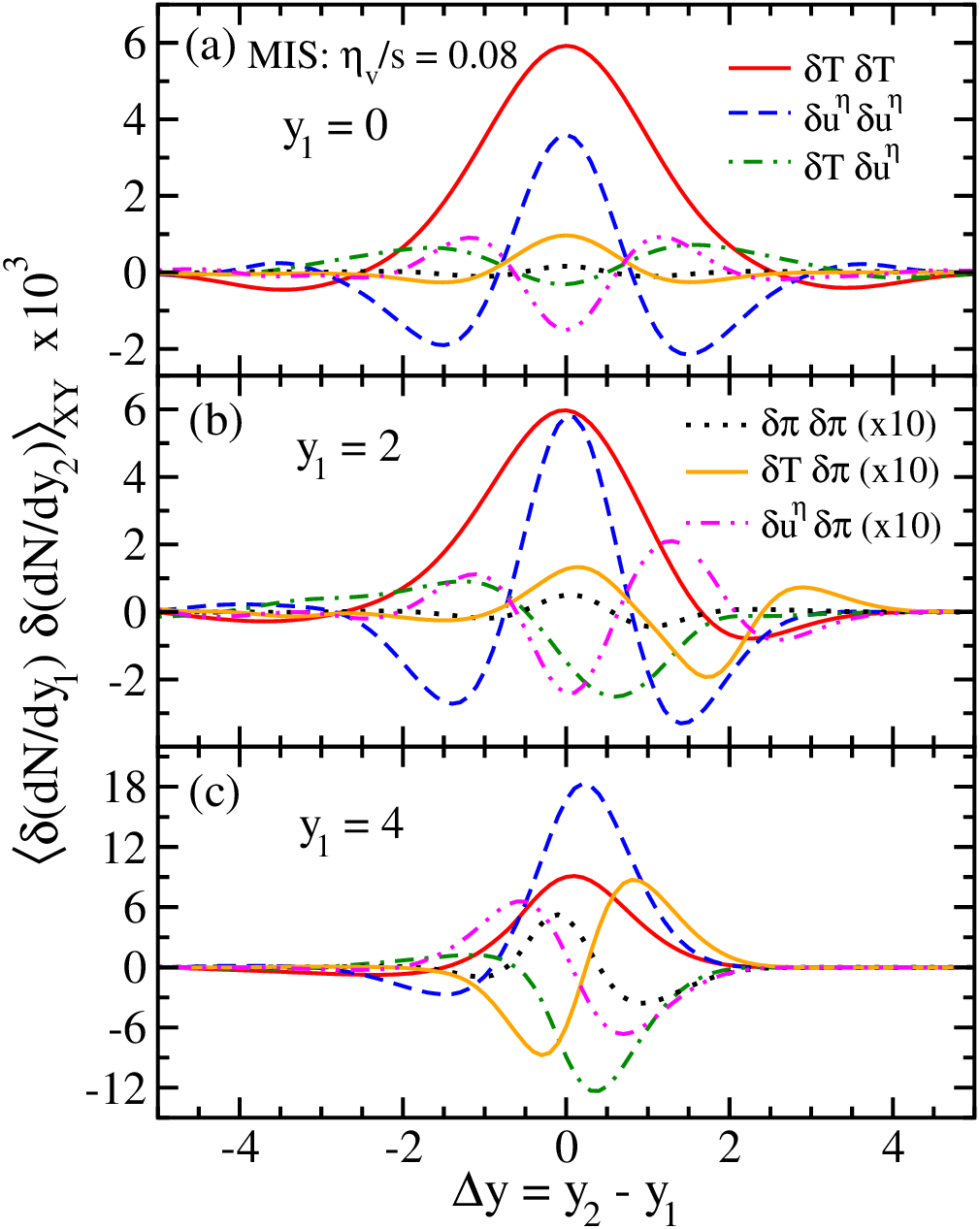}
\caption{Two-particle rapidity correlations for different fluctuations calculated 
for charged pions as a function of pion-rapidity separation $\Delta y = y_1 -y_2$ 
at rapidities $y_1 = 0, 2, 4$ in the MIS hydrodynamics. 
The results are for ideal gas EoS with $\eta_v/s = 1/4\pi$ 
and the initial and final conditions are the same as in Fig. \ref{fig:noiscor}.}
\label{fig:FCor_MIS}
\end{figure}

In this section we will study the effects of thermal fluctuations on two-particle 
rapidity correlations for charged pions in expanding non-boost-invariant 
background fluid. As discussed above, the singularities in the two-point correlators 
$\langle X(\eta_1) Y (\eta_2) \rangle$ (with $X,Y \equiv \delta T, \delta u^\eta, \delta \pi$) 
of Eq. (\ref{FRapCor:eq}) are smeared out by the function ${\cal F}_{X,Y}$ leading to clear 
observable structures in the computed correlations at freeze-out. In Fig. \ref{fig:FCor_MIS}
we present the various rapidity correlators $\langle (\delta dN/dy_1) (\delta dN/dy_2) \rangle_{X,Y}$
for charged pions as a function of kinematic rapidity separation $\Delta y = y_1-y_2$ in the MIS
theory with $\eta_v/s = 0.08$ in the average and noise parts of the evolution equations. 
The correlators get broadened when these are convoluted with the smearing functions ${\cal F}_{\delta T}$ 
(which is roughly Gaussian about $\Delta\eta = 0$) and ${\cal F}_{\delta u^\eta}$ 
(which has peaks at $\Delta\eta \approx 1.5$ and vanishes at $\Delta\eta=0$).

As also evident from Fig. \ref{fig:frz_MICE}, the two-pion rapidity correlations about 
midrapidity $y_1=0$ of a pion (see Fig. \ref{fig:FCor_MIS}(a)) is dominated
by temperature-temperature correlation at $\Delta y =0$. At $\Delta y \gtrsim 2$ the distinct
rapidity dependent structures in the correlations $\delta T \delta T$, $\delta u^\eta \delta u^\eta$ and 
their cross correlations contribute almost equally to the long-range rapidity correlations. 
The correlations associated with the shear stress tensor $\delta\pi$ are found quite small at all
rapidity separations.

At large pion rapidity $y_1 > 0$, inspite of smaller magnitude of initial energy densities and hence
reduced strength of noise source as evident from Eq. (\ref{noiseIS:eq}), the enhanced longitudinal
velocity gradients induce larger fluctuations especially for the velocity correlations. 
Figures \ref{fig:FCor_MIS}(b), (c) show that with increasing pion rapidity, the correlations 
involving $\delta u^\eta$ and  $\delta\pi$ become increasingly important. 
However, the correlations here are short-ranged as the fluctuations produced at large $y_1$
reach the freeze-out hypersurface quickly without substantial spreading.
Moreover, the negative correlations about $\Delta y \sim 0$ become appreciable 
so that the total contribution to the two-pion correlation would be smaller than at midrapidity.

\begin{figure}[t]
\includegraphics[width=\linewidth]{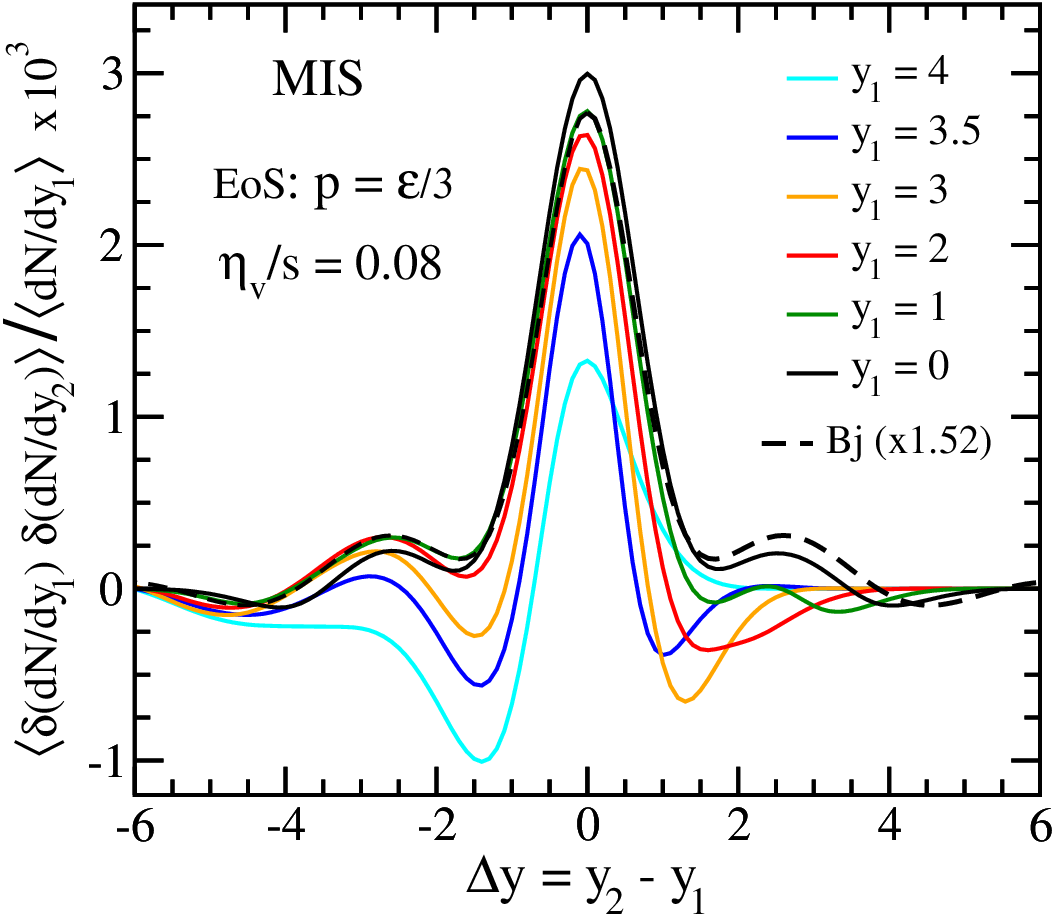}
\caption{Correlation function of charged pions normalized with single-particle rapidity 
distribution in (1+1)D hydrodynamic expansion as a function of rapidity separation 
$\Delta y = y_1 -y_2$ at various rapidities $y_1$.
The results are in the MIS theory for ideal gas EoS with $\eta_v/s = 1/4\pi$ 
and the initial and final conditions are the same as in Fig. \ref{fig:noiscor}.
The corresponding correlation in the Bjorken expansion but normalized by 
$(dN/dy_1)_{y_1=0}$ is shown by black dashed line. }
\label{fig:N12_MIS}
\end{figure}

In Fig. \ref{fig:N12_MIS}  we present the two-particle rapidity correlation for charged pions
at various rapidities $y_1$ in the MIS viscous evolution for an ultra-relativistic
gas EoS with $\eta_v/s = 1/4\pi$ in both the noise and background evolution.  
This has been computed by summing the various components of the noise correlations  
as in Eq. (\ref{FRapCor:eq}) and displayed in Fig. \ref{fig:FCor_MIS}.
For the (1+1)D viscous expansion, the correlations at small rapidities $y_1$ produce pronounced 
short-range peaks and interesting structures at large rapidity separation. On the
other hand, for larger pion rapidities $y_1$ the correlations result in smaller 
peaks at $\Delta y$  and are largely asymmetric about midrapidity. Furthermore, the
singularities mainly from self-correlation at $\Delta y$ are found to be substantial.

We also show the corresponding correlations for boost-invariant expansion (Bjorken flow) in the MIS
theory computed with the same initial time and constant initial energy density as given in Table I 
at $T_{\rm dec} = 150$ MeV. For equivalent comparison the correlation in the Bjorken case 
is normalized by the rapidity density $(dN/dy_1)_{y_1=0}$ for the non-boost-invariant expansion.  
Even after this scaling, the short-range correlation at mid-rapidity is found to be 
slightly larger for the (1+1)D case due to more contribution from self-correlations on 
the freeze-out hypersurface.

\begin{figure}[t]
\includegraphics[width=\linewidth]{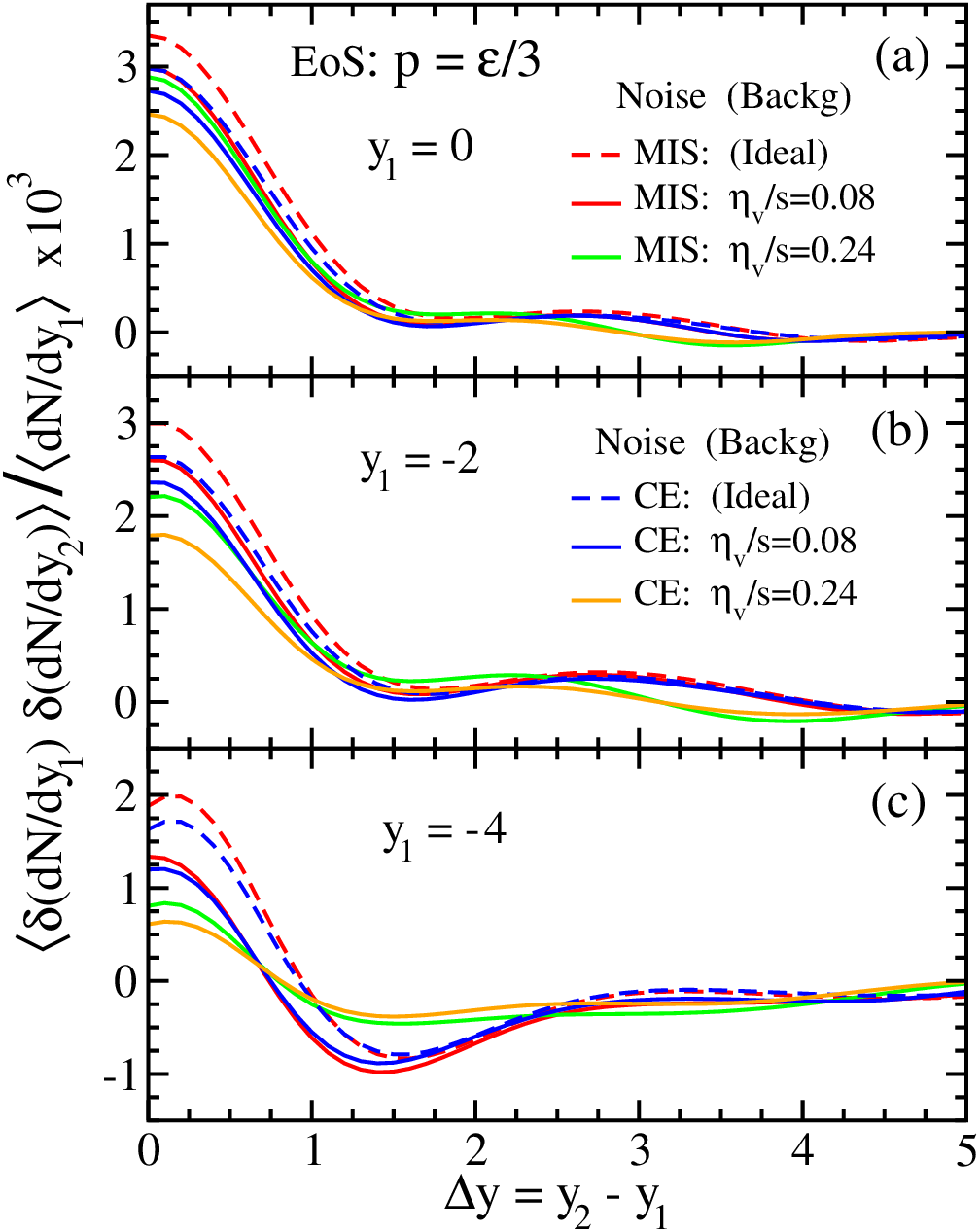}
\caption{Correlation function of charged pions normalized with single-particle rapidity 
distribution as a function of rapidity separation $\Delta y = y_1 -y_2$ at various rapidities $y_1$.
The results are in the  M\"uller-Israel-Stewart (MIS) and Chapman-Enskog (CE) formalisms
for thermal noise evolution and compared with the ideal background hydrodynamic evolution.
An ideal gas EoS ($p=\epsilon/3$) is used and the the initial and freeze-out conditions are given
in Table I.}
\label{fig:N12_MCN}
\end{figure}

Figure  \ref{fig:N12_MCN}(a)-(c) compares the two-particle rapidity correlation for charged pions
in the  M\"uller-Israel-Stewart (MIS) and Chapman-Enskog (CE) dissipative evolutions for 
an ultra-relativistic gas EoS. Using ideal hydrodynamics for the background evolution and 
MIS (red dashed line) and CE (blue dashed line) theories for the evolution of thermal noise 
with $\eta_v/s = 0.08$, we find that for all pion rapidities $y_1$,
the short-range correlation peak at $\Delta y \approx 0$ has a larger magnitude in MIS than 
in CE. This arises due to the smaller damping coefficient $\lambda_\pi$ in MIS 
Eq. (\ref{pievol:eq}) leading to larger 
fluctuations as also evident from Fig. \ref{fig:frz_MICE} for the noise correlators at freeze-out.
At larger $\Delta y$, the singularities in the correlators are more prominent only for large
pion rapidities $|y_1| \sim 4$ resulting in somewhat clear separation of the structures in MIS
and CE formalisms. 

On inclusion of viscosity in the background evolution (solid lines), the correlation 
strengths at $\Delta y$ are suppressed due to viscous damping at small rapidities $y_1$.
However, the long-range structures at large rapidity-separation are rather insensitive 
to viscosity in both the MIS and CE theories. 
Note that the initial energy densities have been readjusted to reproduce the charged hadron
rapidity distribution as given in Table I. 
It is important to note that compared to the Bjorken evolution \cite{Chattopadhyay:2017rgh}, 
in the present non-boost-invariant dynamics the fluctuations cause somewhat
smaller short-range correlation peak ($\Delta y \sim 0$) at larger values 
of particle rapidity $y_1$. 
A larger $\eta_v/s = 0.24$ in the fluctuation evolution leads to further damping of the
correlations due to smearing of the peaks associated with sound horizon.

\begin{figure}[t]
\includegraphics[width=\linewidth]{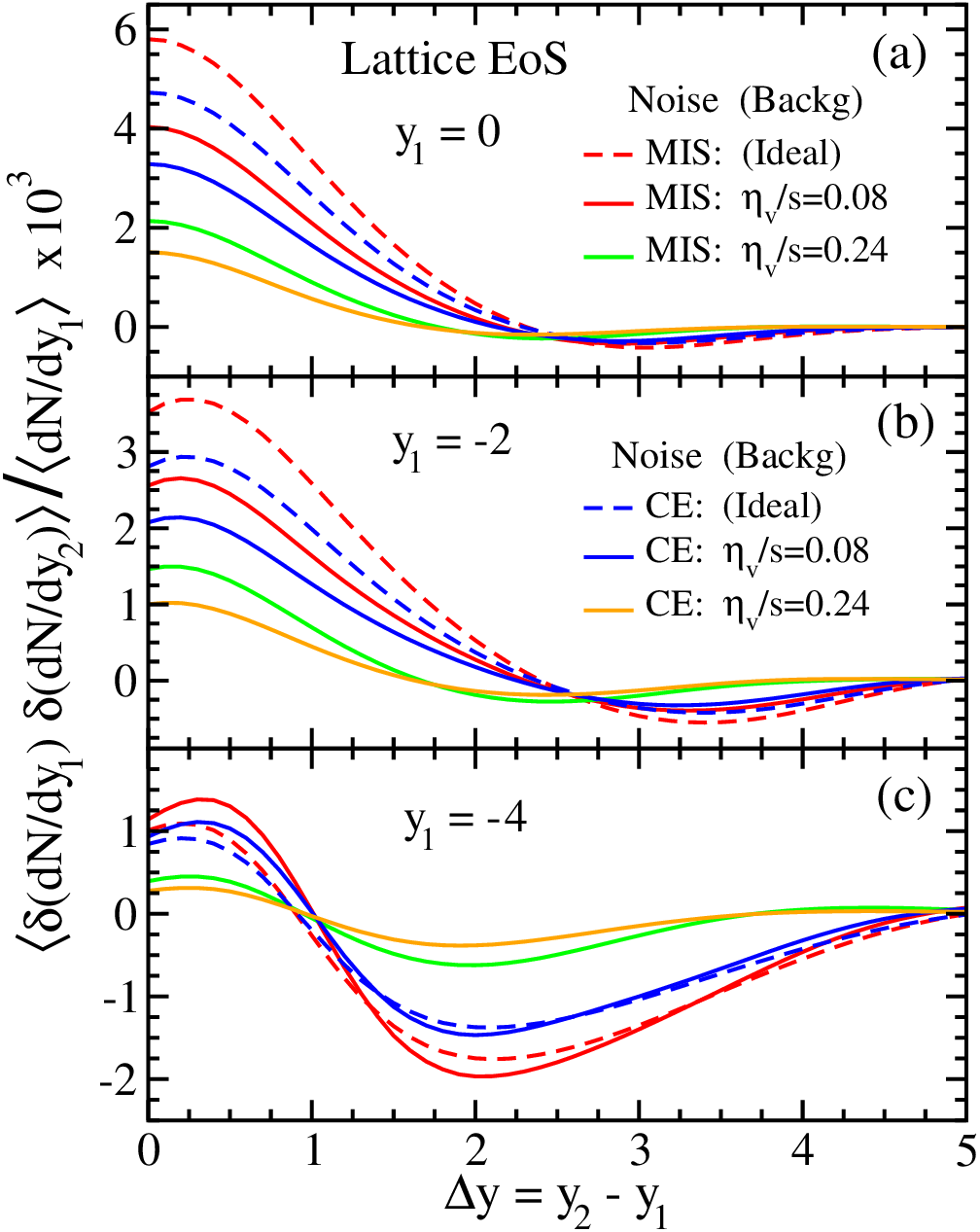}
\caption{Similar to Fig. \ref{fig:N12_MCN} but with a lattice EoS. 
The initial and freeze-out conditions are given in Table I.}
\label{fig:N12_MCN_lat}
\end{figure}

In Fig.  \ref{fig:N12_MCN_lat} we compare the two-particle rapidity correlations for charged pions
in the MIS and CE viscous evolutions but for a lattice QCD EoS. We recall from Table I that the
freeze-out time is somewhat smaller compared to that in the conformal EoS. 
Considerably enhanced two-pion correlation is found at about $\Delta y \sim 0$ for the
lattice QCD EoS as compared to ideal gas EoS, with and without viscosity
in the background evolution. This is primarily due to smaller velocity of sound
in the medium with a lattice EoS that slows down the propagation of fluctuation
over large separations. Here the effects of viscous damping on the rapidity correlations 
is found to be quite significant.

\section{Summary and Conclusions}

We have studied the evolution of thermal noise on top of a non-boost invariant medium 
expansion within the linearized hydrodynamic framework in both MIS and CE dissipative 
formalisms. The (1+1)D equations for the background (averaged) were solved using a newly
developed code based on the SHASTA-FCT algorithm. Using a MacCormack type method to solve 
the linearized perturbation equations, we first studied the correlations induced by a single 
local disturbance propagating on top of the background medium, and then computed two-particle rapidity 
correlations induced by thermal fluctuations which are essentially disturbances (sources) that
persist throughout hydrodynamic expansion. For a single perturbation introduced at some 
space-time rapidity $\eta_0$, the self-correlations induced on the $T_{\mathrm dec}$ hypersurface 
were shown to increase with $\eta_0$, with the velocity-velocity correlator showing the maximum 
growth due to the background acceleration.  Our results for the two-particle correlations show 
that unlike in the Bjorken scenario where correlations depend only on the rapidity separation $\Delta y$,
for the (1+1)D expansion these structures strongly depend on the rapidity $y$ of the final state particle.
Although at $y \sim 0$, the short-ranged structures ($\Delta y \sim 0$) are dominated by the 
temperature-temperature correlations, for large rapidities $y \geq 2$, the velocity-velocity 
correlations are responsible for the self-correlations. 
Inclusion of viscosity was found to reduce the auto-correlations in all the formalisms.
For the lattice QCD EoS with smaller speed of sound, the correlations became larger at small $\Delta y \sim 0$ 
and long-range correlations get reduced, as compared to the ultra-relativistic EoS, due to a lesser extent of 
propagation of fluctuations in the former scenario.

\end{document}